\documentclass{article}
\usepackage[utf8]{inputenc}
\usepackage{authblk}
\usepackage[cm]{fullpage}
\usepackage[reqno,intlimits,sumlimits,centertags]{amsmath}
\usepackage{amssymb,amsfonts,amsthm,amssymb,amsbsy,amscd}
\usepackage{amsfonts}
\usepackage{wrapfig}
\usepackage{graphicx}
\usepackage{courier}
\usepackage{caption}
\usepackage{subcaption}
\usepackage{float}
\usepackage{yfonts}
\usepackage{mathrsfs}
\usepackage{mathtools}
\usepackage{color}
\usepackage[table]{xcolor}
\usepackage{multirow}
\usepackage{tikz}
\usepackage{makecell}
\usepackage{enumitem}
\usepackage{hyperref}
\usepackage[T1]{fontenc}
\usepackage{scalerel,stackengine}
\newtheorem*{theorem*}{Theorem}

\def\taus{s}

\usepackage{todonotes}

\definecolor{colorJRblue}{rgb}{0.,0.,1.}%67}
\definecolor{colorJRred}{rgb}{1.,0.,0.}%67}
\definecolor{colorJRgreen}{rgb}{0.,0.6,0.}%67}

\title{Numerical Continuation and\\ Bifurcation in Nonlinear PDEs:\\ 
Stability, invasion and wavetrains\\ in the Swift-Hohenberg equation}

\author[1]{Ryan Goh}
\affil[1]{Boston University, USA, rgoh@bu.edu}
\author[2]{David Lloyd} 
\affil[2]{University of Surrey, d.lloyd@surrey.ac.uk}
\author[3]{Jens D.M. Rademacher} 
\affil[3]{Universit\"at Hamburg, jens.rademacher@uni-hamburg.de}

\date{}

\begin{document}

	\maketitle

    \tableofcontents

\pagebreak 

\section{Introduction}\label{s:intro}

In this chapter\footnote{This manuscript has been written as a contribution to the planned Handbook on Nonlinear Dynamics. Volume 2 Numerical Methods edited by Vincent Acary.} we continue the topic of numerical continuation and bifurcation from the previous chapter, now with a focus on applications to partial differential equations (PDE). This is a broad topic, and specific issues arise that reflect how different types of PDE require different treatment. Numerical continuation has been, for instance, applied to climate models and three-dimensional fluid problems to help explain the associated complicated dynamics \cite[e.g.]{NetSanchez2015,KUV2012} and the review papers \cite{Dijkstra2014,Dijkstra2019}, and to mechanical contact problems \cite{HPD2000}. We refer to \cite{p2pbook,SanchezNet2016} for a more numerical perspective and further examples. In the short present chapter, we do not aim at a review in any generality, rather our goal is to provide an entry point for interested students and colleagues to the application of continuation methods in pattern formation and coherent structures. An earlier review of numerical continuation in this area can be found in \cite{CS2007} with a focus on the existence problem, while we are more interested in stability.

Although the methods we discuss are much more general, for the sake of clarity, we restrict our attention to a prototypical and seemingly simple semilinear parabolic evolution equation, the cubic Swift-Hohenberg equation (SHE)
\begin{equation}\label{e:SHeqn}
    \partial_t u = F(u,\mu):= -(1+\Delta)^2u + \mu u - u^3,\qquad u(t,x)\in\mathbb{R},\qquad x\in\mathbb{R}^d,
\end{equation}
with a focus on $d=1$ and some aspects concerning $d=2$. Here $\mu\in\mathbb{R}$ is the control parameter, and the differential operators $\partial_t$ and $\Delta = \partial_x^2+\partial_y^2$ for $d=2$, or $\Delta=\partial_x^2$ for $d=1$, denote partial differentiation in time and space, respectively.  In this chapter, we will discuss how numerical continuation methods can be used to study various stability properties of wavetrains, and to compute invasion fronts of wave trains into the unstable zero state. 
While wavetrains are a classical topic (in particular for the SHE \eqref{e:SHeqn}), we include some less well-known aspects that are needed for invasion fronts. This topic connects to the current forefront of research concerning spreading speeds and farfield-core decomposition. We hope that our presentation provides an introductory guideline that can also be used for teaching.

\medskip
Understanding how patterns form is a fundamental problem of active research in many areas of science, including such distant fields as botany \cite{plants} 
%from the emergence of disk florets in daisies 
%{pennybacker2015phyllotaxis} 
and liquid films \cite{thin}. 
%{thiele1998dewetting}. 
Pattern formation is often viewed as an emergent, self-organised phenomenon that results from intrinsic dynamics, not from forcing with a specific spatio-temporal signature and not from boundary effects. Hence, we consider constant coefficients and unbounded domains, thus generating Euclidean symmetry, so that the formation of patterns can also be viewed as spontaneous symmetry breaking. Such symmetry is a marked difference to ordinary differential equations. In this context, pattern formation can in particular occur when a quiescent state is linearly unstable to sinusoidal perturbations leading to the creation of stationary spatially periodic patterns. This has been widely studied in the context of reaction-diffusion systems. 

\begin{figure}[ht]
    \centering
\includegraphics[width=0.6\linewidth]{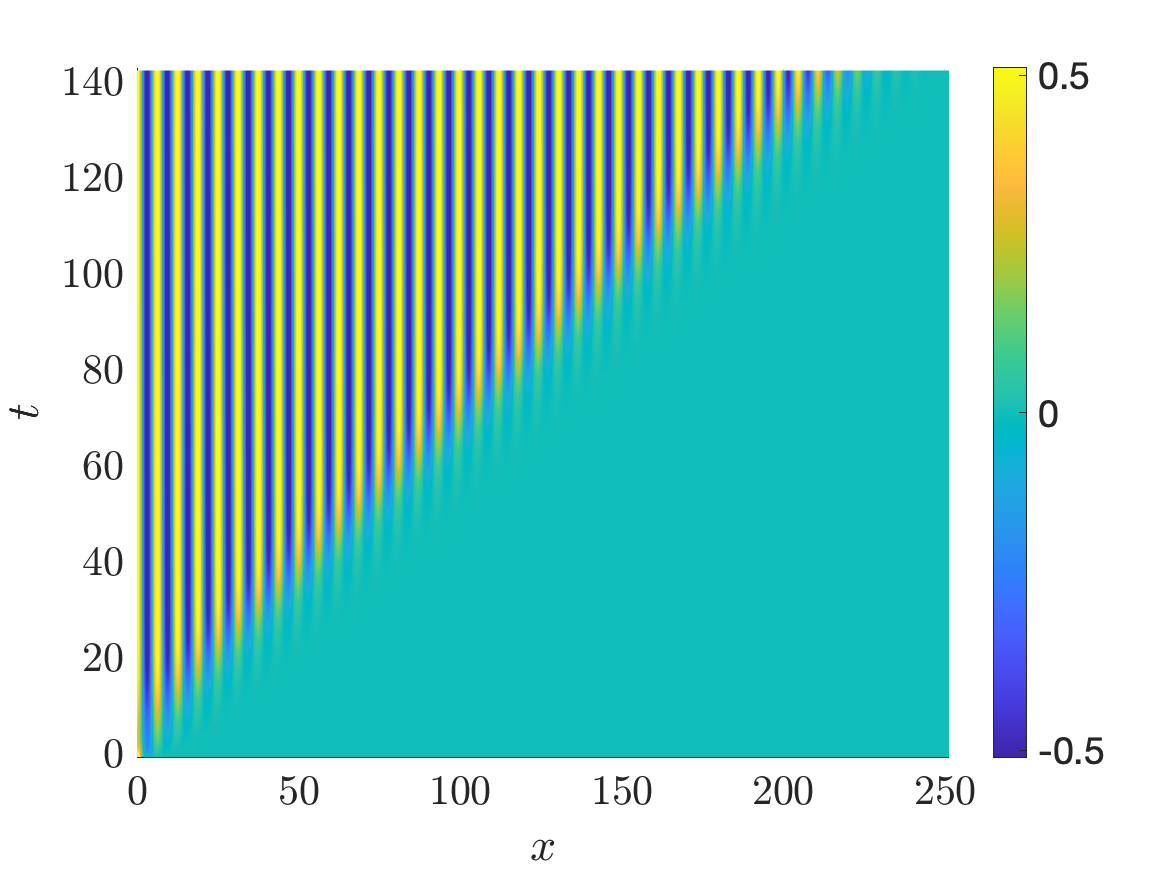}
    \caption{Space-time diagram of pattern-forming invasion front in the Swift-Hohenberg equation \eqref{e:SHeqn} with $\mu = 0.25$ and localized initial condition $u(x,0) = \frac{1}{2} e^{-x^2}$; color coding for $u(x,t)$. The computation was done with 4th order exponential time-differencing and spectral discretization in space. The underlying only partially shown computational domain is $x\in[-80\pi,80\pi]$ with $dt = 0.01$ and $N = 2^{13}$ Fourier modes.    A stable wavetrain (upper left area) invades the homogeneous unstable state, cf.\ Figures~\ref{f:wt-existence} and~\ref{f:SHE-Busse}; time-slices near an invasion front are shown in Figure~\ref{fig:sh-front}.}
    \label{f:SHE-invasion}
\end{figure}

To understand such a process one naturally starts with the existence and stability properties of the patterns, here stationary wavetrains that generically come as a family parameterised by their wavenumber. A more challenging and in general much less understood aspect is %to understand
the actual pattern formation process that results from applying a spatially localised perturbation. 
Indeed, patterns can invade an (unstable) quiescent state in a coherent way via a moving interface that spatially connects the background state and the pattern, here the wavetrain. This interface can move with a fixed (appropriately defined) speed and selects a certain wavelength of the wavetrain in its wake.  See Figure~\ref{f:SHE-invasion}. 
Although invasion of a stable pattern into an unstable state has a long history of research \cite{van2003front} in both experiment and numerical simulation, it appears that the numerical {continuation} of this pattern-selection problem has not been done before. However, closely related recent progress readily admits the formulation of algorithms to do this computation. Herein certain linear stability properties of the  background state and wavetrains are crucial and thus we begin with this. 

\medskip
In pattern formation and the broader study of coherent structures, often simple patterns serve as building blocks for more complicated ones. Indeed, wavetrains are also building blocks for spatially localised pattern formation, related to so-called homoclinic snaking, cf.\ e.g.\ \cite{knobloch2015spatial} and the references therein. Here families of steady states occur, whose spatial profiles asymptote to a homogeneous background state, but are near a wavetrain in a bounded interval. Analogously, in 2D spatially extended hexagon patterns relate to spatially localised patterns and snaking, and the background state need not be homogeneous \cite{bramburger2024localized}. 
In fact, these phenomena can be found in \eqref{e:SHeqn} with modified nonlinearity. 

\medskip
Numerical bifurcation and continuation methods as outlined in this chapter, admit to locate and compute parameterised families of coherent structures without simulating the PDE. As in the previous book chapter, ultimately the algorithms use a root-finding method for nonlinear equations, typically Newton-like, so that much of the preparatory work is to formulate the problem as a well-conditioned root-finding problem for a locally unique solution. 
Developing such algorithms is based in particular on bifurcation analysis, which is analogous to finite dimensional problems. Although the actual computations are always finite dimensional, for PDE problems this dimension can be very large, which requires some optimisation of algorithms. We only touch upon some aspects in \S\ref{s:SHE-fronts} and refer to \cite[e.g.]{SanchezNet2016,AVD2016} and also the classic \cite{AGbook} for a more numerical perspective. 
More specific to PDE with Euclidean symmetry is the presence of continuous spectrum, which, for instance, leads to a multitude of nearly critical eigenvalues. To handle this requires an a priori stability analysis in advance of a numerical formulation, in particular using spatial dynamics, where an unbounded space direction is viewed as the evolution variable. Since this appears less well-known, we focus on it more than on specific numerical methods, for which we also refer to \cite[e.g.]{p2pbook,CS2007}. As mentioned before, we focus on the SHE \eqref{e:SHeqn}, but the methods apply much more broadly, in particular to reaction-diffusion systems. Concerning the discussion of spectra for homogenous states and wavetrains we essentially use the approach presented in \cite{JRSanSch2007}, which is somewhat less accessible due to complications that do not arise in the scalar \eqref{e:SHeqn}. 

\medskip
This chapter is organised as follows. We begin with a discussion of elementary properties of \eqref{e:SHeqn} and the onset of wavetrains via the Ginzburg-Landau formalism in \S\ref{s:SHEbasic}. In \S\ref{s:wtspec} we discuss formulations suitable for numeric continuation of existence and stability of wavetrains extended to striped patterns in 2D. In a similar manner, the more subtle issues of convective and absolute instability, as well as linear spreading speeds are outlined in \S\ref{s:instab}. Based on this, we explain the elements of farfield-core decomposition and numerical continuation for invasion fronts of wavetrains in \S\ref{s:SHE-fronts}. Source codes used to produce many of the computational results depicted throughout this work can be found at the GitHub repository \url{https://github.com/ryan-goh/wavetrains-and-invasion-swift-hohenberg}. 
Here we opted for some variety:  For some basic continuation of wavetrains we use AUTO \cite{auto}; for the stability boundaries and spectra of wavetrains we give an implementation based on the single self-contained secant-continuation routine from \cite{avitabile2020}; for weighted and absolute spectra of the trivial state we provide a MATHEMATICA notebook; for the invasion fronts we give a MATLAB implementation. 
%(or \cite{code_rg_jr_dl}).

\section{Onset of pattern formation}\label{s:SHEbasic}
The SHE \eqref{e:SHeqn} possesses a variational structure via the functional 
\[
\mathcal{P}_\mu(u)=\tfrac12\|(1+\Delta)u\|_2^2-\tfrac12\mu \|u\|_2^2
+\tfrac14 \|u^2\|_2^2,
\]
so that %for $u_0\in L^2(\mathbb{R}^2)$ 
a straightforward formal computation yields the dissipation equation 
\begin{align}\label{e:dissSH}
\frac{d}{dt}\mathcal{P}_\mu(u(t))=-\|F(u(t),\mu)\|_2^2
\quad, t>0, 
\end{align}
or $\frac{\mathrm{d}}{\mathrm {d}t}\|u\|_2^2= - \nabla \mathcal{P}_\mu(u)$. 
This is, for instance, justified with initial data in $L^2(\mathbb{R}^2)$ and useful to study global-in-time properties of solutions. Steady states solve $0=F(u,\mu)$  and given such a state $u_*(x)$ at some $\mu=\mu_*$, the local geometry of $\mathcal{P}$ is relevant for stability properties. These are largely determined by the spectrum of the linearisation $\partial_u F(u_*,\mu_*)$, which is essentially characterised by bounded solutions to the eigenvalue problem 
\[
\lambda w = \partial_u F(u_*,\mu_*) w
\]
with eigenvalue parameter $\lambda\in\mathbb{C}$. For spatially inhomogeneous $u_*$, translation symmetry implies a priori that $\lambda=0$ is a solution to the eigenvalue problem with $w=\partial_x u_*$ or $w=\partial_y u_*$, or both, as can be seen from differentiating $0=F(u_*,\mu_*)$ with respect to $x,y$. Indeed, bounded solutions to the eigenvalue problem correspond to $\lambda$ in, e.g., the $L^2$-spectrum of $\partial_u F(u_*,\mu_*)$. See, e.g., \cite{SandstedStabReview} for a more general context. In this chapter,  $L^2$-spectrum refers to the spectrum of the operator in question viewed as a closed unbounded operator on $L^2(\Omega)$ with $\Omega\subset \mathbb{R}^d$ from the context and domain a Sobolev space  $H^j(\Omega)$ with $j$ the order of the differential operator; in the present formulation $H^4(\mathbb{R}^2)$. 

Specifically, the trivial state $u\equiv 0$ is a solution of \eqref{e:SHeqn} for all $\mu$ and 
since $\mathcal{P}_\mu$ has a global minimum at $u\equiv 0$ for $\mu\leq 0$, the trivial state is globally asymptotically stable in this case.  For local aspects, the linearisation $\partial_u F(0,\mu)$ is given by 
\[
L_0(\mu) := \partial_u F(0,\mu) = \mu -(1+\Delta)^2, 
\]
and the $L^2$-spectrum $\Sigma_\mu(0)$ can be found upon Fourier transforming in $(x,y)$ with wavevector $(\ell_x,\ell_y)\in\mathbb{R}^2$ of wavelength $\ell=\sqrt{\ell_x^2+\ell_y^2}$. This gives 
\[
\widehat L_0(\ell;\mu) = \mu-(1-\ell^2)^2, \quad
\Sigma_\mu(0) = \{\widehat L_0(\ell;\mu): \ell\in\mathbb{R}\}.
\]
Since the spectrum is strictly stable for $\mu<0$, we infer local exponential asymptotic stability (actually global due to the above observation). 
For $\mu>0$ the spectrum and thus the trivial state is unstable and at $\mu=0$ the spectrum at zero stems from the circle of Fourier modes $e^{i(\ell_xx+\ell_yy)}$ with wavenumber $\ell=1$. Such a spectral configuration occurs more broadly in nonlinear equations posed on the plane and is referred to as finite wavelength instability, or Turing instability in reaction-diffusion systems. 
When $\mu$ increases beyond $\mu=0$, the change of stability yields the bifurcation of a multitude of solutions. Next we discuss the simplest type, namely the bifurcation of spatially periodic states that we refer to as \emph{wavetrains}. 

\medskip
We note that there are variants of SHE to which  the approach outlined in this chapter can also be applied rather directly, in particular the `quadratic-cubic' and `cubic-quintic' SHE, where the nonlinearities contain other monomials. Conceptually, the approach presented in this chapter can be applied much more broadly, in particular to reaction diffusion systems and various fluids models.

\subsection{Bifurcation of wavetrains}\label{s:wtonset}

\begin{figure}
\centering{\small    
    \begin{tabular}{cc}
    \includegraphics[width=0.4\linewidth]{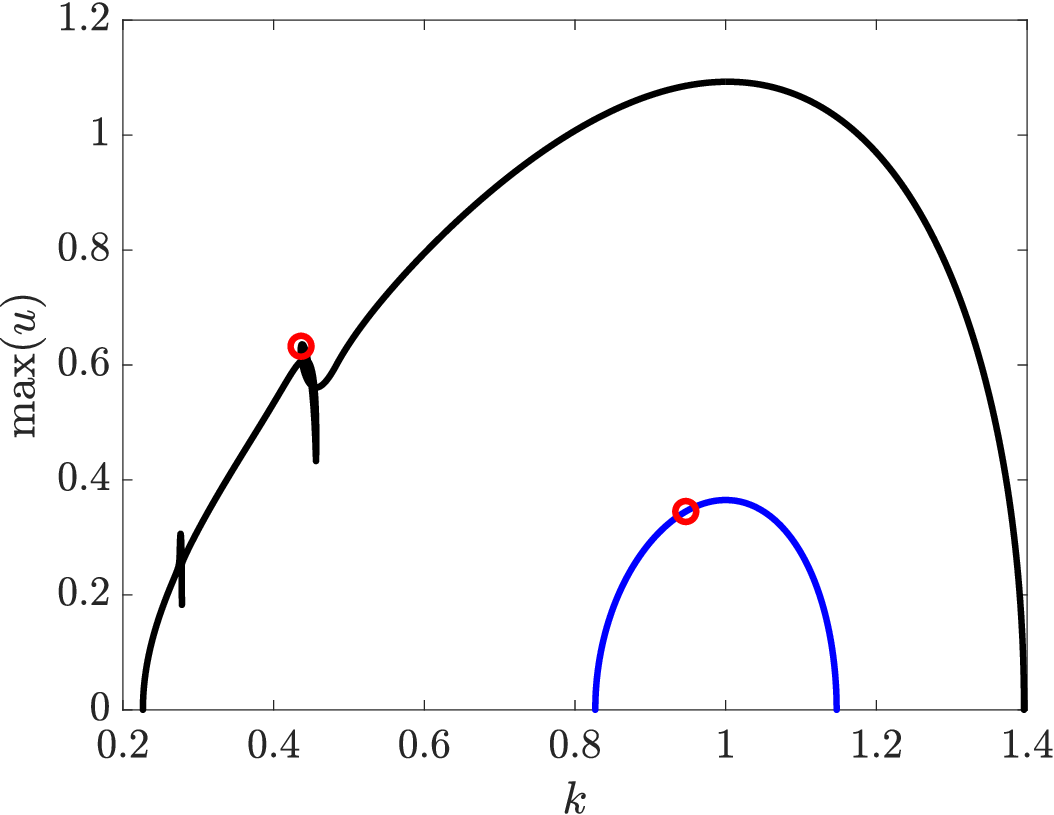}&
    \includegraphics[width=0.4\linewidth]{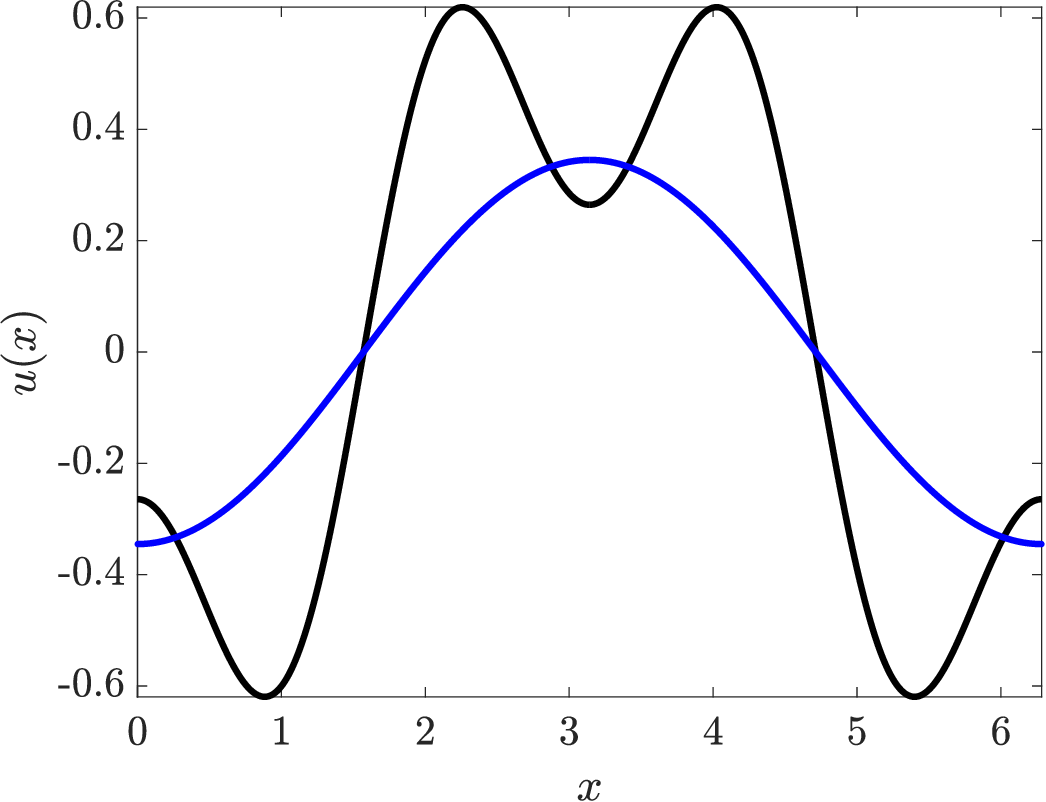}\\
    (a) & (b)
     \end{tabular}
    }
    \caption{Numerical continuation of wavetrains in \eqref{e:SHeqn} implemented in AUTO \cite{auto}. (a) Bifurcation diagram for $\mu=0.1$ (blue) and $\mu=0.9$ (black) in terms of the wavenumber $k$; 
    (b) solution profiles at marked locations in (a). }
    \label{f:wt-existence}
\end{figure}

The emergence of spatially periodic solutions as $\mu$ increases beyond zero can be anticipated from the aforementioned Fourier modes associated to $\lambda=0$ as the point in the spectrum of $\Sigma_0(0)$ with maximal real part. In particular, $e^{\pm i x}$ are among these modes so that perturbations by $\cos(x)$ are among the first to grow as $\mu$ becomes positive. As can be seen from the dissipation equation, this growth is bounded, and it turns out that it settles in steady states. (For general perturbations this is not clear.) 

Let us reduce to one space dimension in the remainder of this subsection, so that  $\Delta=\partial_x^2$. It is well known \cite[e.g.]{Mielke1997} and will be discussed further below, that the bifurcation at $\mu=0$ gives rise to small amplitude stationary, spatially $2\pi/k$-periodic states of the form $u(x,t) = u_p(kx)$, i.e., wavetrains, and we refer to $k$ as their (nonlinear) wavenumber. 
These wavetrains are even and $2\pi$-periodic in the phase variable $\xi=kx$, and one can show that 
\begin{align}\label{e:wavetrain}
u_p(\xi) = u_p(\xi;\mu,k)=\sqrt{\tfrac1 3(\mu-4\kappa^2)}\,\cos(\xi) + \mathcal{O}(|\mu-4\kappa^2|^{3/2}),
\end{align} 
%\JR{I think the formula above is not correct since it does not match with the Ginzburg-Landau analysis. It should be: 
%$u_p(\xi) = u_p(\xi;\mu,k)=\sqrt{\tfrac1 3(\mu-4\kappa^2)}\,\cos(\xi) + \mathcal{O}(|\mu-4\kappa^2|^{3/2})
%$. There seems to be a typo in Mielke's paper. 
%}
%\DL{I concur\ldots the GZ analysis gives what you have written down and there is a typo in Mielke's paper}
where $\mu\in(4\kappa^2,\mu_0]$ 
with $\kappa=k-1$. 
Hence, for each $0<\mu\ll1$ there is a family of wavetrains parameterised by their wavenumber $k$ in terms of $\kappa$, with $|\kappa|<\sqrt{\mu}$. See the blue curves in Figure~\ref{f:wt-existence}. 
The cubic SHE \eqref{e:SHeqn} features the additional odd symmetry $F(u,\mu)=-F(-u,\mu)$, which implies the additional symmetry $u_p(x+\pi)=-u_p(x)$. 

To understand the bifurcations of the wavetrains $u_p$, let us consider the ODE in $x$ for steady states of SHE given by $F(u,\mu)=0$; note  $F(u_p(k\cdot),\mu)=0$ and recall that we here set $\Delta=\partial_x^2$. This ODE is reversible due to the reflection symmetry $x\to -x$ (another reversibility stems from the additional reflection symmetry $u\to-u$), and it has Hamiltonian structure: In terms of the variables
\[
q_1=u,\quad q_2=\partial_x u, 
\quad p_1=-(\partial_x u+\partial_x^3u), \quad p_2 = u+\partial_x^2u,
\]
and defining the potential 
$P_\mu(q) = \frac \mu 2 q^2 - \frac 1 4 q^4$, 
this ODE can be written as
\begin{align}
\dot p_1 = p_2 - P_\mu'(q_1), \quad
\dot p_2 &=-p_1, \quad
\dot q_1=q_2, \quad
\dot q_2=p_2-q_1.
%    \dot q_1&=q_2  & \dot p_1 &= p_2 - P_\mu'(q_1)\\
%    \dot q_2&=p_2-q_1 & \dot p_2 &=-p_1, 
\label{e:SHspatial}
\end{align}
which is Hamiltonian with respect to
\[
H_\mu(p_1,p_2,q_1,q_2) = p_1q_2-p_2q_1+ \tfrac 12 p_2^2+P_\mu(q_1).
\]
In particular, for $\mu>0$ the potential $P_\mu$ possesses a local minimum at $q=u=0$. The reversibility is here in terms of the reflection $(p_1,p_2,q_1,q_2)\to (-p_1,p_2,q_1,-q_2)$. The linearisation of this system at zero (which corresponds to $L_0(\mu) u=0$) has eigenvalues $\pm\sqrt{1\pm\sqrt{\mu}}$. For $\mu<0$ all of these have nonzero real part, for $\mu=0$ they form a  complex conjugate pair $\pm i$ of multiplicity two, and for $\mu>0$ they form two distinct pairs of complex conjugate eigenvalues on the imaginary axis. This is the eigenvalue signature of a Hamiltonian-Hopf bifurcation, which can here be seen as a reversible 1:1 resonant Hopf-bifurcation. It is well known that this gives rise to a family of reversibly symmetric periodic orbits that `encircle' the trivial solution for $\mu>0$. These are parameterised by their period over an interval whose endpoints are the imaginary parts of the eigenvalues at the selected $\mu>0$, at which amplitude tends to zero \cite[e.g.]{Knobloch1994}. See also \cite[Sec. 4.3]{haragus2011local}.

The specific expansion \eqref{e:wavetrain} can be derived from the proof, but more easily from the following well known (at first formal) computation based on a suitable scaling ansatz in terms of $\mu\approx 0$. It is useful to perform this more generally for the time-dependent one-dimensional SHE \eqref{e:SHeqn} through an amplitude modulation ansatz 
\begin{align}\label{e:SHansatz}
u(t,x) = \varepsilon A(T,X)e^{ix} + c.c. + h.o.t
\end{align}
with the parabolic scaling $X=\varepsilon x$, $T=\varepsilon^2 t$. In addition, we link this to $\mu$ via  $\mu=\varepsilon^2\tilde\mu$ where  $\tilde\mu\in\mathbb{R}$ is redundant but useful. 
Substituting this into SHE and sorting by powers of $\varepsilon$ gives \cite[e.g.]{SchneiderUeckerBook}, at order $\varepsilon^3$, the so-called real Ginzburg-Landau equation
\begin{align}\label{e:rGL}
\partial_T A = 4\partial_X^2 A + \tilde\mu A -3|A|^2A.
\end{align}
It possesses the gauge symmetry that  $e^{i\varphi}A$ is a solution if $A$ is,  for any $\varphi\in\mathbb{R}$. This explains the occurrence of stationary wavetrain solutions of the plane wave form
\[
A_p(X) = A_p(X;\tilde\mu,\tilde\kappa) = r e^{i \tilde\kappa X}, \; r=\sqrt{\tfrac13(\tilde\mu-4\tilde\kappa^2)},
\]
for $\tilde\mu>4\tilde \kappa^2$. Substituting this into the ansatz \eqref{e:SHansatz} gives the leading order form of $u_p$ in \eqref{e:wavetrain} for $\tilde\mu=1$. 

For the actual numerical computation of $u_p$ in \eqref{e:SHeqn} by continuation, which relies on a Newton method, we need to formulate the existence problem of wavetrains suitably to ensure local uniqueness. Wavetrains $u_p(\xi)$ are solutions to \eqref{e:SHeqn} for $\xi=kx$ that are $2\pi$-periodic in $\xi$ -- in general in a co-moving frame $z=\xi-ct$. We thus seek solutions to the ODE boundary value problem 
\begin{equation}\label{e:wtBVP}
\begin{aligned}
     -c\partial_z u_p &= -(1+k^2\partial_z^2)^2u_p+\mu u_p - u_p^3, \; z\in (0,2\pi)\\
     \partial_z^ju_p(2\pi)&=\partial_z^j u_p(0), \; j=1,\ldots,4.
\end{aligned}
\end{equation}
Due to reversible symmetry for standing wavetrains, where $c=0$, we could replace the periodic boundary conditions by homogeneous Neumann-type on half the domain $z\in[0,\pi]$. However, for later purposes, numerical stability and generality, it is useful to stick to periodic boundary conditions, although this means that solutions to \eqref{e:wtBVP} are not locally unique due to translation symmetry. Abstractly, local uniqueness of $u_p$ on a solution branch parameterised by $\mu$ can be  ensured by constraining to functions that are $L^2$-orthogonal to translations of $u_p$ via $\langle \partial_z u_p, \partial_\mu u_p\rangle_{L^2} =0$. 
This is simplified by a finite difference approximation of $\partial_\mu u_p$ from a given evaluation $u_p(\cdot;\mu_{\rm old})$ of $u_p$ at a previous continuation step %\JRx{Refer to chapter 3 in this book?} 
with parameter $\mu_{\rm old}$, which gives $\langle \partial_z u_p(\cdot;\mu_{\rm old}), u_p(\cdot;\mu_{\rm old})-u_p(\cdot;\mu)\rangle_{L^2} =0$. Integration by parts with periodic boundary conditions shows $\langle \partial_z u_p(\cdot;\mu_{\rm old}), u_p(\cdot;\mu_{\rm old})\rangle_{L^2} =0$ so that we obtain the simplified so-called phase condition 
\begin{equation}\label{e:wt_phase}
\langle \partial_z u_p(\cdot;\mu_{\rm old}), u_p\rangle_{L^2} = \int_0^{2\pi} \partial_\xi u_p(z;\mu_{\rm old}) u_p(z;\mu) dz.
\end{equation}
It can be readily checked that the combination of \eqref{e:wtBVP} and \eqref{e:wt_phase} generically has invertible linearisation, which suffices for the continuation method \cite{Champneys2007}. 

For larger values of $\mu$ various forms wavetrains emerge and the geometry as well as bifurcation structure becomes rather complicated. See the black curves in Figure~\ref{f:wt-existence}. One aspect is that for $\mu>1$ a pitchfork bifurcation of steady states occurs that generates steady solutions of \eqref{e:SHeqn} that are spatially heteroclinic orbits between these. In turn, these heteroclinics can be connected with the family of wavetrains in $(k,\mu)$-parameter space. 
Concerning the complexity of the set of solutions we refer to \cite{JB2008} and the references therein.

\subsection{Stability in the real Ginzburg-Landau equation}\label{s:stabGL}
Concerning stability of $u_p$, it can be shown that stability properties of $A_p$ in  \eqref{e:rGL} give the correct prediction for small $|\mu|$ via \eqref{e:SHansatz}. Let us therefore briefly consider the stability of $A_p$, which can be done largely explicitly in contrast to that for $u_p$, which will be discussed in \S\ref{s:wtspec}. For this it is useful to consider $A$ and the complex conjugate $\bar{A}$ as independent variables in the system formed by \eqref{e:rGL} and the conjugate equation for $\bar A$. This allows to linearize $|A|^2A = A^2 \bar A$ in $A_p$, but the resulting operator contains the periodic term $A_p^2(X)$, which prevents direct access to the spectrum. However, the gauge symmetry allows to remove this by switching to detuned variables $a=Ae^{-i\tilde\kappa X}$ and its complex conjugate. This amounts to changing $\partial_X$ to $\partial_X+i\tilde\kappa$ and $A_p$ turns into the constant state $a_p=r$. The linearisation in these variables has \emph{constant coefficients} so that Fourier transform gives a block-diagonal linearisation. The block for wave number $\ell\in\mathbb{R}$ is given by the matrix
\begin{align}
\widehat{B}_p(\ell;\tilde\kappa,\tilde\mu):=
\begin{pmatrix}
\tilde\mu-6r^2-4(\ell+\tilde\kappa)^2 & -3r^2\\
-3r^2 & \tilde\mu-6r^2-4(\ell-\tilde\kappa)^2
\end{pmatrix}.
\end{align}
Its characteristic equation for eigenvalue $\lambda\in\mathbb{C}$ gives the linear dispersion relation and $L^2$-spectrum
\[
d_{\rm wt}(\lambda,\ell)=\det\left(\widehat{B}_p(\ell;\tilde\kappa,\tilde\mu)-\lambda\mathrm{Id}\right), \;
\Sigma(A_p)= \{\lambda : \exists \ell\in\mathbb{R},\,d_{\rm wt}(\lambda,\ell)=0\}.
\]
Due to the algebraic structure, the spectrum is the union of two smooth curves parameterised by $\ell$ and possibly branch points at double roots of $d_{\rm wt}(\cdot,\ell)$. Much about the dispersion relation can be understood analytically, in particular the stability boundary as explained next. However, it is worth pointing out that one can use this formulation to compute these curves by numerical continuation with respect to the parameter $k$.

As a consequence of translation symmetry we have  $d_{\rm wt}(0,0)=0$, i.e., $0\in\Sigma(A_p)$, so that the spectrum near $\lambda=0$ is expected to be a smooth curve $\lambda_0(\ell)$ parameterised by $\ell\approx0$. Upon implicit differentiation we readily find that $\lambda_0'(0)=0$ and 
\begin{equation}\label{e:GLEckhaus}
\lambda_0''(0)= \frac{2}{r^2}(\tilde\mu-12\tilde\kappa^2),
\end{equation}
so that the parameterisation of the spectrum is indeed always smooth at $\lambda=0$ ($r>0$ for wavetrains). Moreover, this part of the spectrum extends into the unstable half plane as $|\tilde\kappa|$ increases above $\sqrt{\tilde\mu/12}$, which means that the wavetrains destabilise when compressed or stretched too much compared to the basic wavenumber. It can be readily shown that wavetrains are indeed spectrally stable in the so-called Eckhaus region $\{\tilde\mu>12\tilde\kappa\}$. %See Figure \ref{f:GL-Eckhaus}. 
(The factor $12$ stems from the coefficients in \eqref{e:rGL}; for unit coefficients it is the commonly known $3$.) 
The onset of instability when crossing the Eckhaus boundary is through infinite wavelength  modes $\ell\approx 0$, which is commonly referred to as Eckhaus- or sideband instability. This occurs in pure form on the idealised unbounded domain  $x\in\mathbb{R}$ so that it cannot be precisely studied on bounded domains with periodic boundary conditions. Indeed, in such a case the critical $\mu$ is shifted to a larger value and the onset is via the longest possible mode in the given interval \cite{BarkleyTuckerman}. In both cases, one typically observes in numerical simulations that perturbations of an unstable wavetrain act to compress or dilate the pattern in $x$ to a stable wavelength, possibly via a `phase slip'.

This scenario for the nature of instability, its impact on the dynamics for nearby solutions, and the geometry of the stability region gives the correct prediction for wavetrains in \eqref{e:SHeqn} near onset $0<\mu\ll1$. The stability region of wavetrains in wavenumber-parameter space is often referred to as the Busse-balloon, although this term was coined for problems in fluids in higher space-dimension $x\in\mathbb{R}^d$, $d>1$, where additional instability mechanims are prominent. See Figure~\ref{f:SHE-Busse}. However,  for \eqref{e:SHeqn} and other equations that feature wavetrains but do not possess gauge symmetry,  stability can (in general) not be determined through an algebraic equation. One then has to determine the critical  spectrum of a linear operator with periodic coefficients. Even in one space dimension, where this boils down to studying the fundamental matrix solution, there is no general explicit analytical approach. Instead, in the next section, we formulate the stability problem in a form that can be directly used for numerical continuation.

\section{Spectrum and stability regions of wavetrains in 2D}\label{s:wtspec}

\begin{figure}
\centering{\small    
    \begin{tabular}{cc}
    \includegraphics[width=0.45\linewidth]{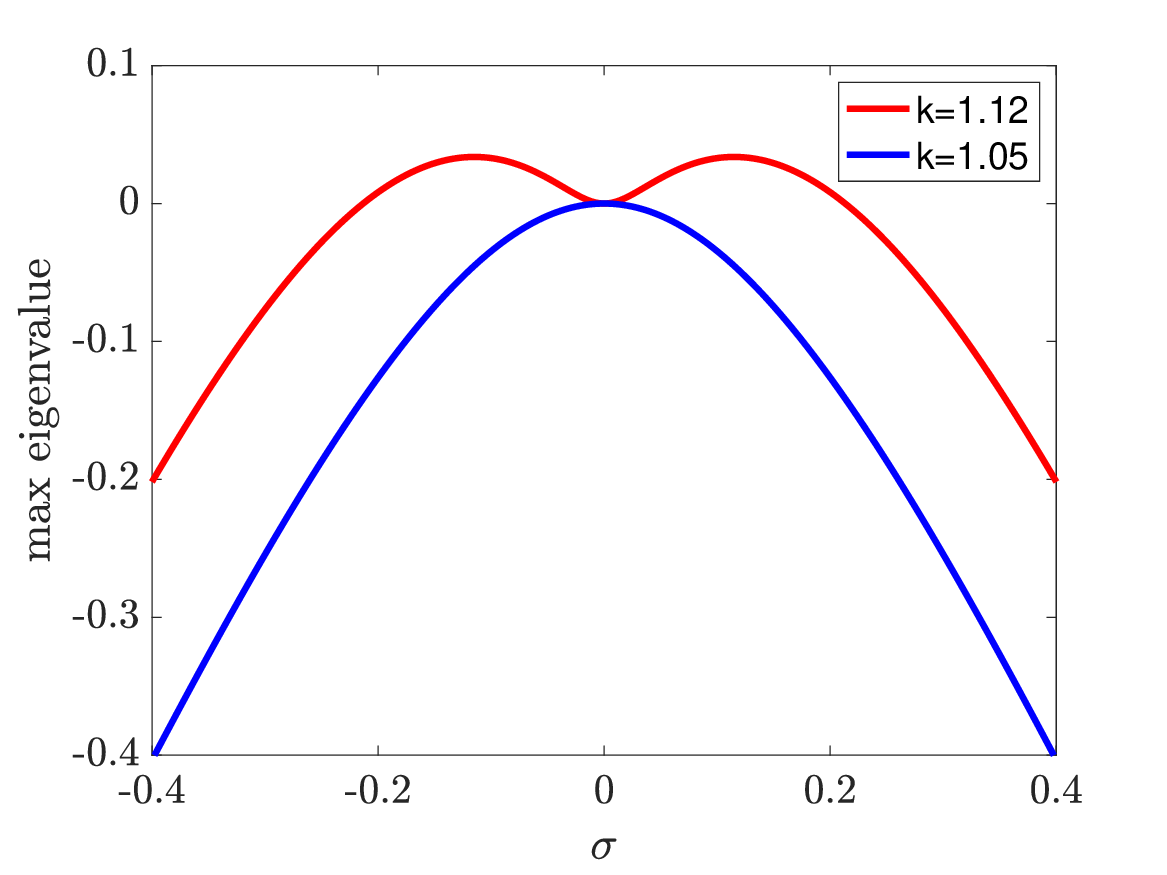}&
    \includegraphics[width=0.45\linewidth]{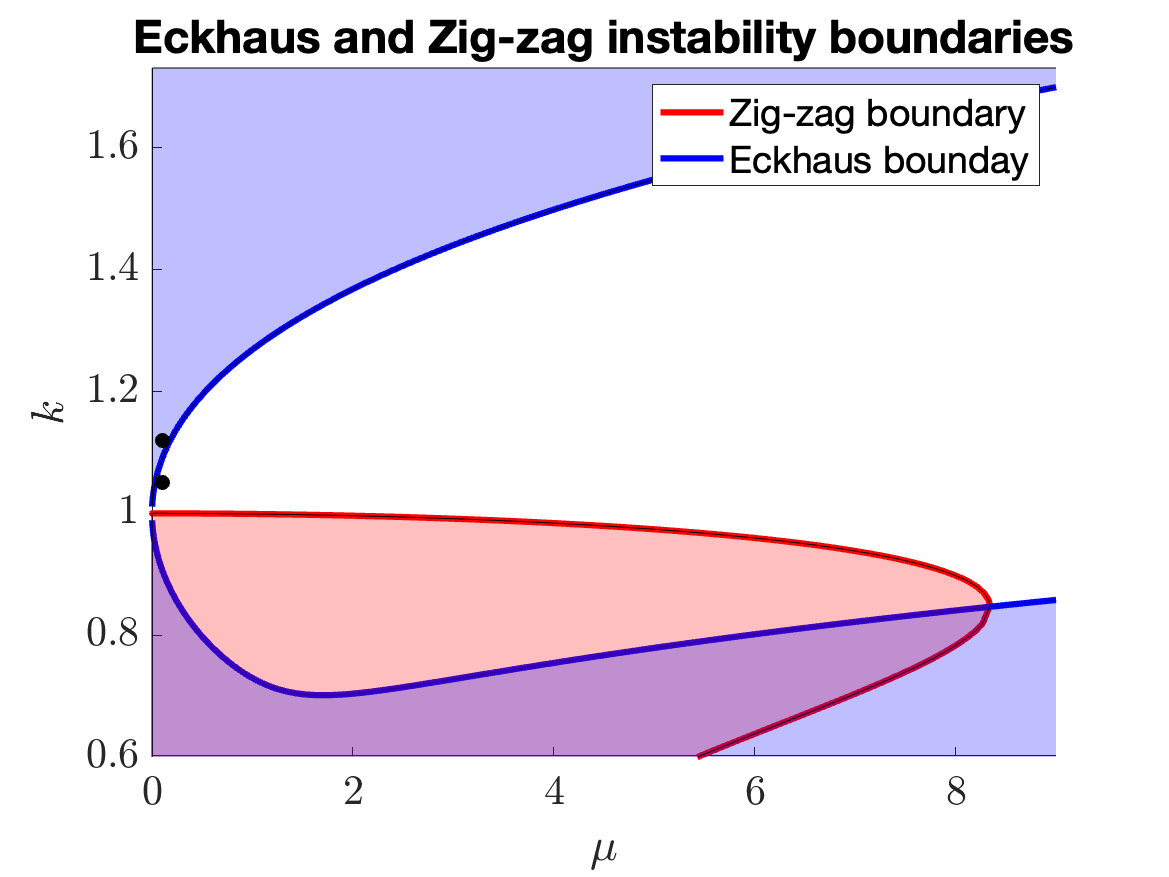}\\%EK_ZZ_boundaries.eps}\\
    (a) & (b)
     \end{tabular}
    }
    \caption{(a) Floquet-Bloch spectrum near $\sigma=0$ for fixed $\mu=0.1$ and one value of $k$ in the stable region, another beyond the Eckhaus boundary as marked by black bullets in (b).  (b) Stability boundaries with unstable region shaded. For larger $\mu$ the $k$-values of the Eckhaus boundary increase further, in particular beyond $k=1$. 
    %The zigzag boundaries for smaller $k$ shows interesting oscillations.
    }
    \label{f:SHE-Busse}
\end{figure}

\noindent
One can trivially extend the one-dimensional $2\pi/k$-periodic pattern, $u_p(kx)$, to the plane by letting $u(x,y) = u_p(kx)$ leading to so-called striped solutions of \eqref{e:SHeqn}. 
It is useful to rescale space $\xi=kx$, which amounts to replacing $\partial_x$ in \eqref{e:SHeqn} by $k\partial_\xi$. In order to analyse the stability we linearise the rescaled \eqref{e:SHeqn} about $u_p(\xi)$ posed on the plane, leading to the linear PDE
\begin{equation}\label{e:wt-lin2D}
    \partial_t w = L(u_p)w,\; L(u_p):=-(1+k^2\partial_\xi^2+\partial_y^2)^2 + \mu  - 3(u_p(\xi))^2,
\end{equation}
for $w=w(\xi,y)$. Here, $u_p(\xi)^2$, and thus the operator $L(u_p)$,  is  $\pi$-periodic in $\xi$ since $u_p(x+\pi) = -u_p(x)$ from the odd symmetry of the cubic nonlinearity; more generally the operator would be $2\pi$-periodic. To study the spectrum $\Sigma(u_p)$ of $L(u_p)$ on the plane $(\xi,y)\in\mathbb{R}^2$, as a first step we can use that the terms of $L(u_p)$ are independent of $y$ and perform a Fourier transform in $y$ with wavenumber $\tau\in\mathbb{R}$. 
This defines the transformed operator $\widehat{L}(u_p;\tau^2)$ which results from $L(u_p)$ by replacing $\partial_y^2$ with $-\tau^2$. With $\Sigma_{\tau}(u_p)$ denoting its spectrum on $L^2(\mathbb{R})$ with respect to the $\xi$-variable we have 
\[
\Sigma(u_p) = \cup_{\tau\geq 0} \Sigma_\tau(u_p)
\]
so that it remains to characterise $\Sigma_\tau(u_p)$. 
As mentioned in \S\ref{s:stabGL}, the eigenvalue problem 
\begin{align}\label{e:evalprob1}
\lambda w = \widehat{L}(u_p;\tau)w,
\end{align}
with $w$ a function of $\xi$ only, can be viewed as a four dimensional $\lambda$-dependent ODE with $\pi$-periodic coefficients. The spectrum $\Sigma_{\mathrm{F},\tau}(\lambda)$ of the resulting $\lambda$-dependent fundamental matrix solution over time $\pi$ is characterised by Floquet exponents $\nu_j(\lambda)$, $j=1,\ldots,4$. In terms of these, \eqref{e:evalprob1} possesses a bounded solution precisely when $\nu_j(\lambda)\in i\mathbb{R}$ for some $j$. Indeed, such $\lambda$ lie in the $L^2$-spectrum and one can show that 
\[
\Sigma_\tau(u_p) = \{ \lambda \in\mathbb{C} : \Sigma_{\mathrm{F},\tau}(\lambda)\cap i\mathbb{R}\neq \emptyset\}. 
\]
The problem to determine $\Sigma_\tau(u_p)$ can thus be translated into determining  purely imaginary Floquet exponents. However, it is generally numerically costly to compute these by solving the characteristic equation of a numerically approximated fundamental matrix solution.

Towards numerical stability and for the general context, it is instructive to take  a more operator theoretic perspective. A useful method motivated from physics for operators with lattice symmetries is to consider Bloch waves. In one space-dimension for $\pi$-periodic coefficients this amounts to the so-called Floquet-Bloch decomposition of the relevant spaces $H^j(\mathbb{R})$, $j\in \mathbb{N}_0$, 
\[
H^j(\mathbb{R}) \simeq \bigoplus_{\sigma\in [0, 2)} %\pi)} 
H^j_\mathrm{per}(0,\pi).
\]
This is given by the isomorphism
\[
w(\xi) = \int_0^{\pi} e^{i\sigma \xi}\tilde w(\xi;\sigma)d\sigma,
\]
where $w\in H^j(\mathbb{R})$ and $\tilde w(\cdot;\sigma)\in H^j_\mathrm{per}(0,\pi)$, which is the $H^j$-space on $[0,\pi]$ based on periodic functions, in particular $\tilde w(\xi;\sigma) = \tilde w(\xi+\pi;\sigma)$. See \cite[e.g.]{Mielke1997} and the review \cite{SandstedStabReview}. 
Operators with periodic coefficients posed on $\mathbb{R}$ can be accordingly decomposed into the so-called Floquet-Bloch operators posed on the bounded intervals $(0,\pi)$ with periodic boundary conditions. Here $\partial_\xi$ is replaced by $\partial_\xi+i\sigma$ with so-called Floquet-Bloch wavenumber $\sigma\in[0, 2)$  %\pi)$. 
or equivalently by rotation symmetry $\sigma\in[-1,1)$. 
In case of $\widehat{L}(u_p;\tau)$ this gives the Floquet-Bloch operators 
\begin{equation}\label{e:FBop}
B(\sigma,\tau;\mu,k):= -(1-\tau^2+k^2(\partial_\xi+i\sigma)^2)^2+\mu-3u_p^2.
\end{equation}
Accordingly, for the spectrum $\Sigma_\tau(u_p)$ of $\widehat{L}(u_p;\tau)$ this implies the decomposition 
\[
\Sigma_\tau(u_p) = \bigcup_{\sigma\in[0,2)}%\pi)} 
\Sigma_{\sigma,\tau}(u_p)
= \bigcup_{\sigma\in[-1,1]}%\pi)} 
\Sigma_{\sigma,\tau}(u_p),
\]
where $\Sigma_{\sigma,\tau}(u_p)$ is the spectrum of $B(\sigma,\tau;\mu,k)$. 
Note the analogy to the computation of the spectrum in the case of the Ginzburg-Landau equation in \S\ref{s:wtonset} and the matrix operators $\widehat{B}_p(\ell;\tilde\kappa,\tilde\mu)$. 

Summarising, we now analyse the stability of a wavetrain $u_p(kx)$ of \eqref{e:SHeqn} via Floquet-Bloch decomposition of perturbations $w(x,y)$, i.e., by writing 
\[
w(x,y) = e^{i(\sigma kx + \tau y)}\tilde w(\xi),
\]
where $\xi=kx$ and $\tilde w\in H^4_\mathrm{per}(0,\pi)$. The eigenvalue problem for $B(\sigma,\tau;\mu,k)$ is then the 4th-order ODE boundary value problem
\begin{equation}\label{e:FB-BVP}
    \lambda \tilde w = B(\sigma,\tau;\mu,k)\tilde w, \, 
\tfrac{d^j}{d\xi^j}\tilde w(0) = \tfrac{d^j}{d\xi^j}\tilde w(\pi),\; j=0,1,\ldots,4.
    %:= -(1+(1+\kappa)(\partial_x + i\sigma))^2 - \tau^2)^2\tilde w +\mu \tilde w - 3u_p^2\tilde w
\end{equation}
for non-zero solutions, which we make unique by adding the normalisation condition 
\begin{equation}\label{e:FB-normalise}
\int_0^{\pi}\tilde w^2 dx = 1.
\end{equation}
Note that the change of variables to $\varphi(x) = e^{i\sigma kx}\tilde w(\xi)$  relates \eqref{e:FB-BVP} to the aforementioned purely imaginary Floquet-exponent of the time-$\pi$-map of the linear ODE. 

\begin{figure}
\centering
\begin{tabular}{cc}
  \includegraphics[width=0.35\linewidth]{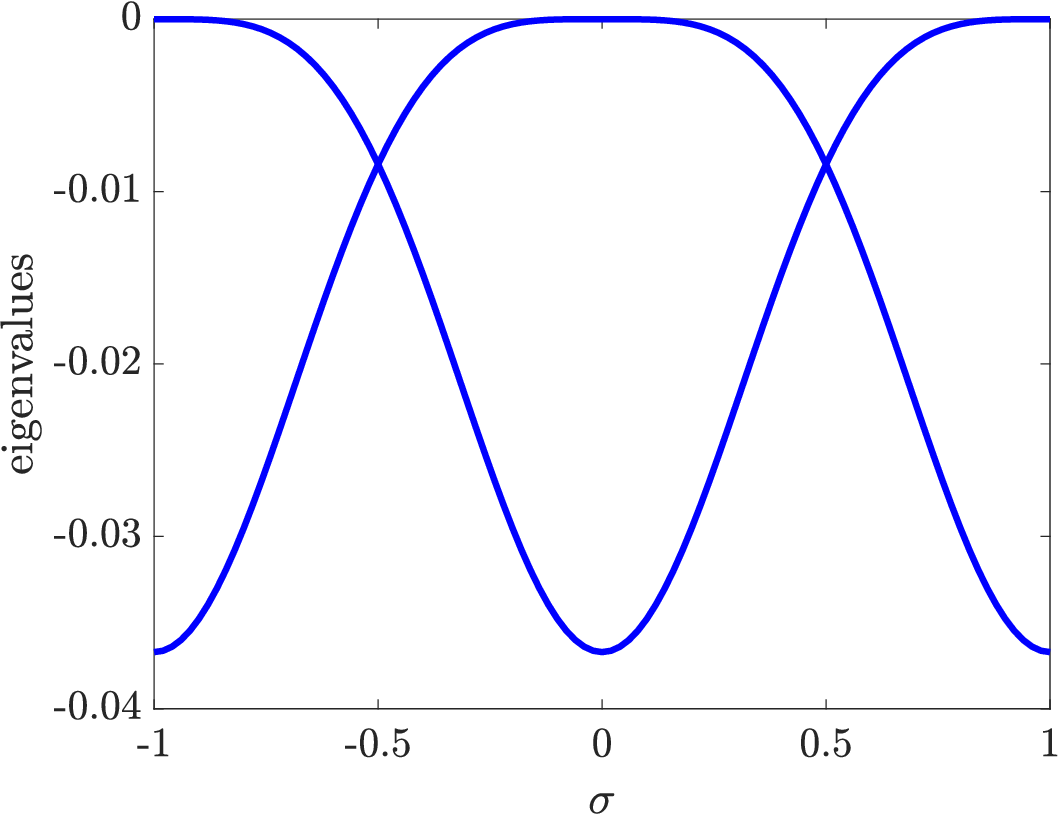}
&  \includegraphics[width=0.35\linewidth]{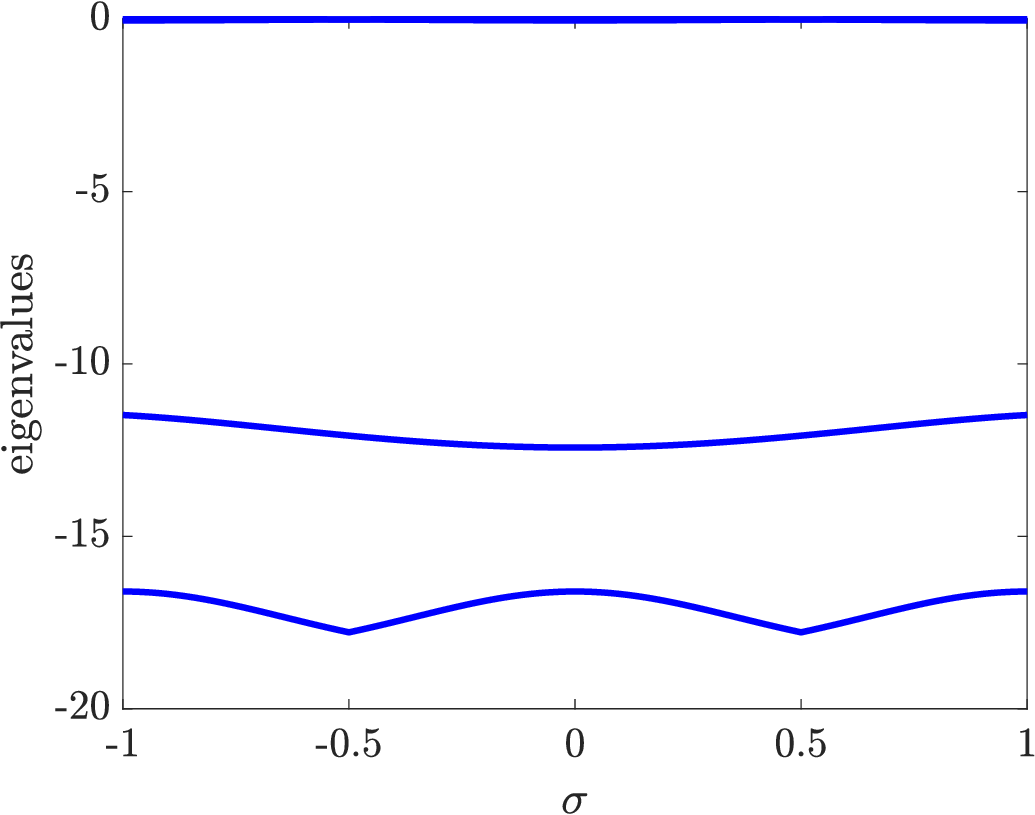}\\
(a) & (b)\\
 \includegraphics[width=0.35\linewidth]{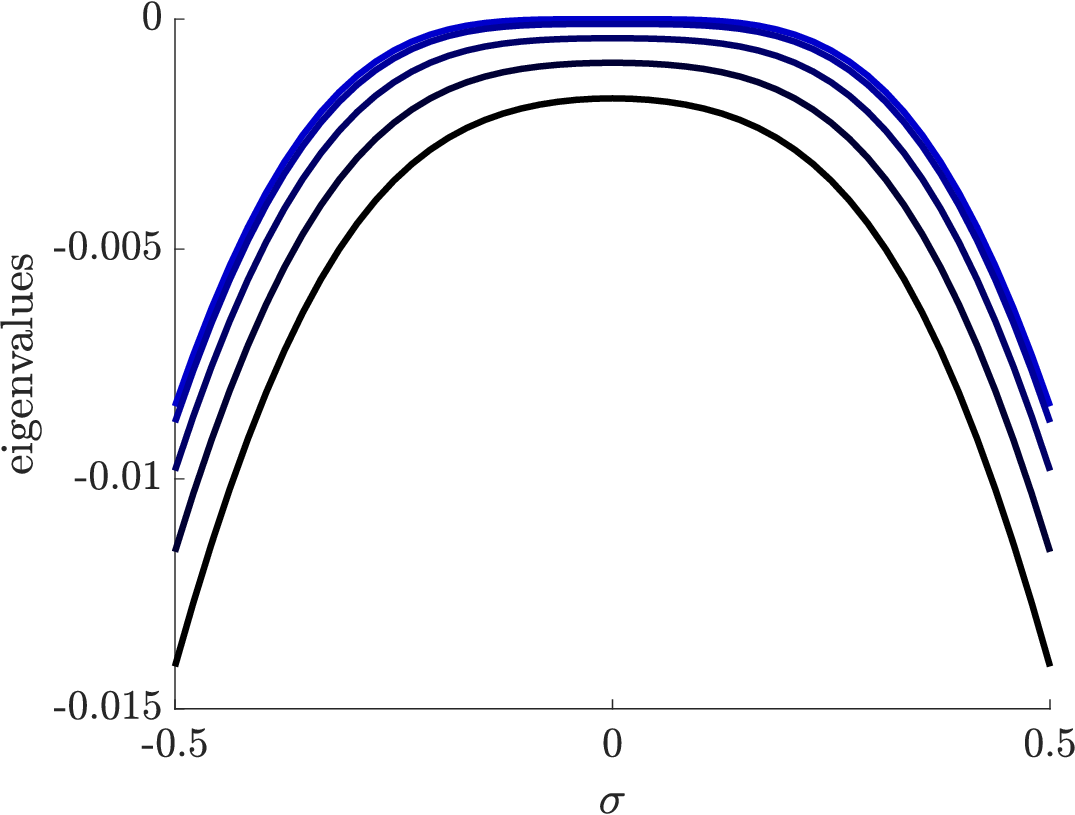}
& \includegraphics[width=0.35\linewidth]{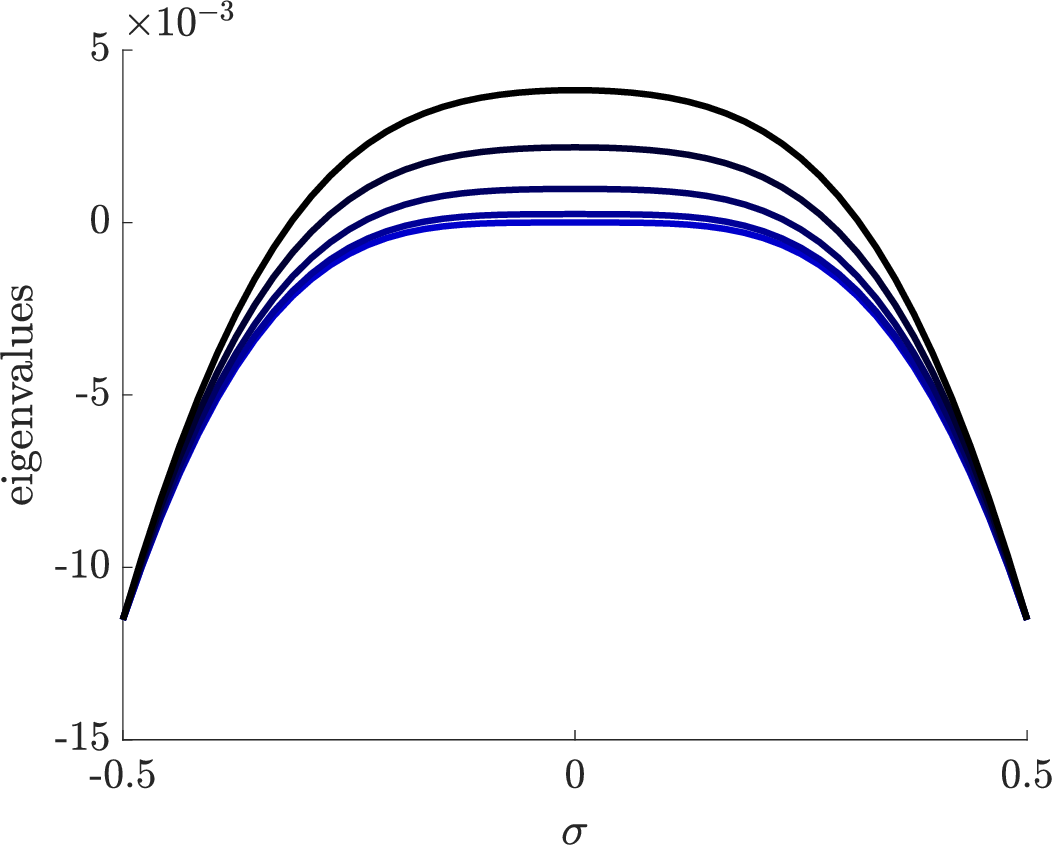}\\
 (c) & (d)
\end{tabular}
    \caption{Parts of the essential spectrum of selected $u_p(k;\mu)$ computed via Floquet-Bloch operator. The direct computation via an eigenvalue solver is in this case faster than the computation by continuation in $\sigma$. Parameters lie on the Eckhaus boundary: For panels (a,b,c)  $\mu\approx 10$, $k\approx 0.9$ just outside the range in Figure~\ref{f:SHE-Busse}; for panel (d) $\mu\approx 5$, $k\approx 0.8$ in the zigzag unstable regime. 
 In panels (a,b) $\sigma\in[-1,1]$, $\tau=0$, and we plot the two, respectively four, most unstable eigenvalues, illustrating the band structure of the spectrum with disconnected intervals. 
 In panels (c,d) we plot only the most unstable eigenvalue for  $\sigma\in[-0.5,0.5]$ and $\tau=j/10$ for $j=1,\ldots 50$.}\label{f:spec}
\end{figure}

The formulation \eqref{e:FB-BVP}, \eqref{e:FB-normalise} can be viewed as the implicit dispersion relation for $u_p$, analogous to $d_{\rm wt}(\lambda,\ell)=0$ in \S\ref{s:stabGL}. It is the basis for the deriving the analytical conditions and algorithms for  computations of spectra and stability boundaries by numerical continuation in the following. In order to illustrate the spectra one obtains, we plot samples in Figure~\ref{f:spec}.

Recall that translation symmetry a priori yields $\lambda=0\in \Sigma(L(u_p))$ with eigenmode $\partial_\xi u_p$. This appears in the Floquet-Bloch operator at $\tau=0$ (since $\partial_\xi u_p$ is constant in $y$) and at $\sigma=0$ (since $\partial_\xi u_p$ is $\pi$-periodic in $\xi$).  Analogous to \S\ref{s:wtonset} we expect that the spectrum near $\lambda=0$ is a smooth surface $\lambda_0$ parameterised by $(\sigma,\tau)\in [0,\pi)\times\mathbb{R}$. 
One can show that $\lambda_0$ determines the stability at least for $|\mu|$ small and the onset of instability of striped solutions in \eqref{e:SHeqn} on the plane comes in two types: The Eckhaus %(aka sideband) 
instability as discussed earlier, and the so-called Zigzag instability \cite{Horst1971,Eckhaus1965,Mielke1997}. Recall the Eckhaus instability occurs when the stripe solution's wavelength is either too short or too long, i.e., it stems from positive curvature in $\lambda_0$ as a function of $\sigma$ at $\sigma=\tau=0$, and the instability acts to compress or dilate the pattern in $x$ to a stable wavelength. See Figure~\ref{f:SHE-Busse}(a). 
The zigzag instability is a long wavelength instability in the transverse $y$ direction leading  to growing $\cos(\ell y)$-type perturbations for large $\ell$, i.e., it stems from positive curvature in $\lambda_0$ as a function of $\tau$ at $\tau=\sigma=0$. See Figure~\ref{f:spec}(d). In particular, both Eckhaus and zigzag instabilities appear in pure form on the unbounded domain only. 

\subsection{Eckhaus stability boundary}\label{s:sidebandnum}

Analogous to \eqref{e:GLEckhaus}, to identify the Eckhaus instability we need access to the curvature coefficient of the parameterisation by $\sigma$ at $\tau=0$ (since $\tau=0$ for $\lambda_0=0$). 
Hence, we view $\tilde w$ as dependent on $\sigma$ and differentiate \eqref{e:FB-BVP} as well as \eqref{e:FB-normalise} twice with respect to $\sigma$, thus deriving equations for  $\partial_\sigma^j \tilde w$,  $\partial_\sigma^j \lambda_0$, $j=1,2$ evaluated at $\sigma=\tau=0$. 
The Eckhaus stability boundary is then characterized by $\partial_\sigma^2\lambda=0$. 
A factor $i$ is convenient for odd derivatives so we define $\tilde w_\sigma:=i\partial_\sigma \tilde w$, $\tilde w_{\sigma\sigma}:=\partial_\sigma^2 \tilde w$,  $\lambda_{\sigma}:=i\partial_\sigma \lambda_0$, $\lambda_{\sigma\sigma}:=\partial_\sigma^2 \lambda_0$, all evaluated at $\sigma=\tau=0$. Since the Eckhaus boundary will be a curve in the $(k,\mu)$-plane it is useful for its computation to solve the resulting equations jointly with the existence problem for wavetrains \eqref{e:wtBVP}, \eqref{e:wt_phase}. Since the existence problem is posed on $(0,2\pi)$ with periodic boundary, for simplicity we extend the Floquet-Bloch domain to this as well. For the calculation of the derivatives with respect to $\sigma$, we note that $B(0,0;\mu,k)=L(u_p)$, and that the evaluations at $\tau=\sigma=0$, where $\lambda_0=0$, lead to real valued problems. 
With this notation and written in terms of $z=\xi-ct$, the combination of existence problem and zeroth up to second derivative of \eqref{e:FB-BVP} (at $\sigma=\tau=0$) reads 
\begin{subequations}\label{e:eckhaus_eqns}
    \begin{align}
        -(1+k^2\partial_z^2)^2u_p + \mu u_p - u_p^3 + c\partial_z u_p =& 0,\label{e:wt-ex}\\
        L(u_p)\tilde w - \lambda_0 \tilde w =& 0,\\
        L(u_p)\tilde w_{\sigma} - \lambda_\sigma\tilde w - 4(1+k^2\partial_z^2){ k^2 }\partial_z\tilde w =& 0,\\
        L(u_p)\tilde w_{\sigma\sigma} + 2\lambda_{\sigma}\tilde w_{\sigma} + 8(1 + k^2\partial_z^2){ k^2 }\partial_z\tilde w_{\sigma} + (4{ k^2 } - \lambda_{\sigma\sigma})\tilde w + 12k^4\partial_z^2\tilde w =& 0.
    \end{align}
\end{subequations}
Similarly, differentiating the phase and normalisation conditions gives
\begin{subequations}\label{e:eckhaus_phase_cond}
    \begin{align}
        \int_0^{2\pi}u_z^{\rm ref}(u_p-u_p^{\rm ref})\mathrm{d}z = 0,\\
        \int_0^{2\pi}\tilde w^2 \mathrm{d}z = 1,\qquad
        \int_0^{2\pi}\tilde w\tilde w_{\sigma}\mathrm{d}z= 0,\qquad \int_0^{2\pi}\tilde w\tilde w_{\sigma\sigma}\mathrm{d}z = 0,
    \end{align}
\end{subequations}
Here we have modified the phase condition \eqref{e:wt_phase} so as to prevent translation relative to a reference profile $u^{\rm ref}(z) = \cos(z)$, which is sometimes numerically more stable.

By allowing for non-zero curvature  $\partial_\sigma^2 |_{\sigma=\tau=0}\lambda_0$ and non-zero $\lambda_0$ this formulation can also be used when seeking marginal stability of other parts of the spectrum, not related to $\lambda_0$. For example striped solutions can undergo a finite wavelength transverse instability. It is also convenient to generate initial conditions for a continuation of the Eckhaus boundary: 

First, given $u_p$ and $\partial_\xi u_p$, for fixed $\mu,k$ sufficiently close to the Eckhaus boundary, use the initial `guess' $\lambda_0=0$,  $\tilde w=\partial_\xi u_p$, $\tilde w_\sigma=\tilde w_{\sigma\sigma}=0$ for a Newton loop of \eqref{e:eckhaus_eqns}, \eqref{e:eckhaus_phase_cond} with variables 
$(u,\tilde w, \tilde w_\sigma, \tilde w_{\sigma\sigma},\lambda_0,\lambda_\sigma,\lambda_{\sigma\sigma},c)$.

Second, fix $\lambda_{\sigma\sigma}=0$ and add $k$ to the variables for another Newton loop that brings the variables onto the Eckhaus boundary. (This requires a good choice of initial $\mu,k$.)

Third, to trace out the Eckhaus stability boundary, add $\mu$ to the variables, thus solving \eqref{e:eckhaus_eqns} with \eqref{e:eckhaus_phase_cond} for $(\mu, k, u, \tilde w, \tilde w_{\sigma}, \tilde w_{\sigma\sigma}, \lambda_0,\lambda_{\sigma},c)$. This (generically) gives a curve in the $(k,\mu)$-plane, selecting the wavetrains $u_p(\xi;\mu,k)$ for which the spectrum is marginally stable in terms of the sideband. 

\medskip
The entire problem can be discretised using, e.g., finite-differences or spectral methods and can be embedded in basically any numerical continuation framework. 
%The code used for the figures of this chapter can be found in the github link in \S\ref{ch1} above. %\cite{code_rg_jr_dl}.)}
We plot the resulting Eckhaus stability boundary in Figure~\ref{f:SHE-Busse}. The underlying  implementation can be found under the github link in \S\ref{s:intro} above.

The cautious reader will notice that for $\lambda_0=0$, \eqref{e:eckhaus_eqns}(b), means $\tilde w=C \partial_z u_p$ with $C= 1/\|\partial_z u_p\|_2$ to satisfy the first equation of \eqref{e:eckhaus_phase_cond}(b). In the Eckhaus boundary computation one could thus remove \eqref{e:eckhaus_eqns}(b) by replacing $\tilde w$ with $C \partial_z u_p$ without loss. However, it is convenient to be able to switch from Eckhaus boundary computation to a continuation for a different purpose.

 \subsection{Zigzag stability boundary}\label{s:zigzagnum}

For the zigzag instability, we set $\sigma =0$ and consider $\tilde w$, $\lambda_0$ as a function of $\tau^2$; note that $B$ in \eqref{e:FBop} depends on $\tau$ only through $\tau^2$. 
Hence, for the curvature information, we only need to differentiate \eqref{e:FB-BVP} once in $\tau =0$ to $\tau^2$, i.e., we need to resolve the terms $\tilde w_i,\lambda_{0i}$, $i=1,2$, where 
\[
\tilde w(x) = \tilde w_0(x) + \tau^2\tilde w_2(x) + \mathcal{O}(\tau^4),\qquad \lambda = \lambda_{00} + \tau^2\lambda_{02} +\mathcal{O}(\tau^4),
\]
which are real as for the sideband problem. Since \eqref{e:FB-BVP} at $\tau=\sigma=0$ is the 1D eigenvalue problem of $u_p$, the relevant solution is $(\lambda_{00},\tilde w_0) = (0,\partial_\xi u_p)$; recall the translation symmetry in $x$. Differentiating \eqref{e:FB-BVP} with respect to $\tau^2$ and evaluating at $\tau=\sigma=\lambda=0$ gives 
\begin{equation}\label{e:zigzag}
\lambda_{02} \tilde w_0 =  2(1+k^2 \partial_\xi^2)\tilde w_0 + L(u_p) \tilde w_2,
\end{equation}
with periodic boundary conditions for $\tilde w_2$. Taking the inner product with $\tilde w_0$, and using that $L(u_p)$ is self-adjoint with kernel $\tilde w_0$, yields
\begin{equation}
    \lambda_{02} = 2\frac{\langle (k^2\partial_\xi^2{ +}1)\partial_\xi u_p,\partial_\xi u_p\rangle}{\langle \partial_\xi u_p,\partial_\xi u_p\rangle}.
%    \lambda_2 = 2\frac{\langle (\partial_x^2-1)(u_p)_x,(u_p)_x\rangle}{\langle (u_p)_x,(u_p)_x\rangle}.
\end{equation}
%\JRx{I get a plus sign, where David had a minus?}
Hence if $\lambda_{02}<0$, i.e. $\lambda_{zz}:=\langle (k^2\partial_\xi^2{ +}1)\partial_\xi u_p,\partial_\xi u_p\rangle<0$, then the striped solutions are zigzag stable and unstable for the opposite sign. Setting $\lambda_{02}=0$, one can readily implement \eqref{e:zigzag} in a continuation code and then trace out the zigzag stability boundary as a function of $\mu$ analogous to the Eckhaus boundary. The resulting zizag stability boundary is also plotted in Figure~\ref{f:SHE-Busse}. The underlying  implementation can be found under the github link in \S\ref{s:intro} above. 

%%%%%%%%%%%%%%%%%%%%%%%%%%%%%%%%%%%%%%%%%%%%%%%%%%%%%
\section{Convective and absolute instability}\label{s:instab}

\begin{figure}
\centering
  \includegraphics[width=0.4\linewidth]{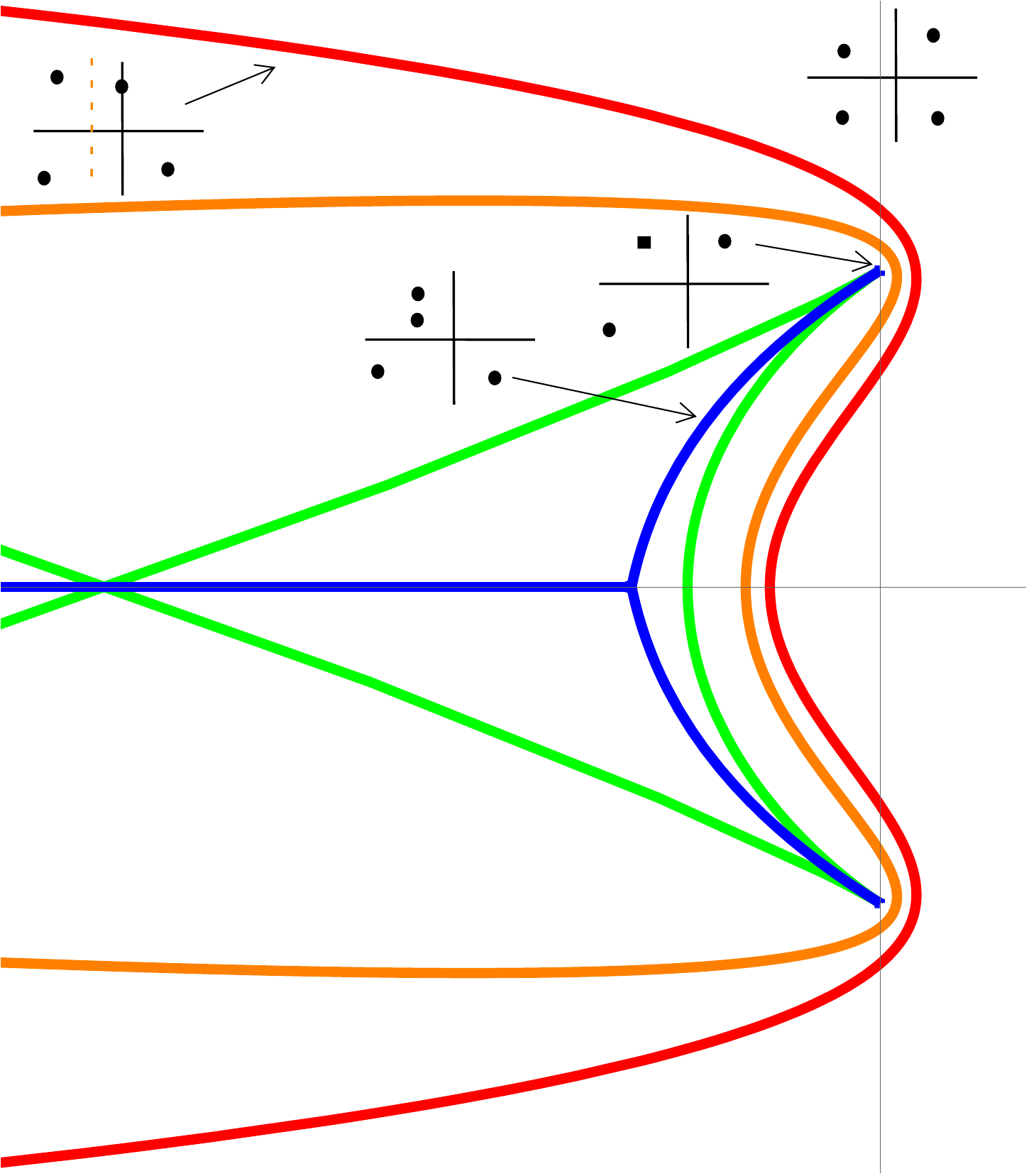}
    \caption{The above figure depicts the unweighted essential spectrum $\Sigma_{\mu,c_*}$ (red), the weighted essential spectrum $\Sigma_{\mu,c_*,\eta}$  for weight with $ \mathrm{Re}\, \nu_*(c_*)<\eta<0$ (orange) and with $\eta = \mathrm{Re}\,\nu_*(c_*)$ (green), and absolute spectrum $\Sigma_{\mu,c_*}^\mathrm{abs}(0)$ (blue), defined below, of the homogeneous state $u = 0$ at the linear spreading speed $c = c_*$ and $\mu = 1/4$. Insets plot the spatial eigenvalues $\{\nu_j\}$ for $\lambda$ values, with square dots denoting a double root. The dashed line in upper left inset denotes the shift caused by weight $\eta$ in orange case. 
    }
\end{figure}

We motivate the upcoming material by a brief heuristic elaboration of the stability of the base state and for simplicity again restrict the domain to real line $x\in\mathbb{R}$. 
Having discussed spectra and stability of wavetrains and homogeneous steady states in $L^2(\mathbb{R})$-based spaces, we now turn to more refined stability properties that capture the interaction of growth and transport.  In particular, we are interested in distinguishing pointwise growth of perturbations from growth in norm that is not pointwise due to transport. The former is referred to as \emph{absolute instability} and the latter as \emph{convective instability}, which can be viewed as a form of stability. We aim at distinctions of these instability types based on spectral properties of the linearisation in the underlying state. This cannot work in spaces with translation invariant norms such as $L^2(\mathbb{R})$, and it is natural to consider weighted norms such that spatially distant quantities appear small. 
As a sidenote, on bounded domains, where boundaries are for some reason absorbing a transported perturbation, the spectrum of the linearisation in a convectively unstable state would predict stability. Indeed, the distinction of absolute and convective stability also relates to spectra on bounded domains, cf.\ \cite{SanSchabs}, but we will not discuss this further here. 

In order to determine the speed of transport in convective instabilities, we consider a frame moving with constant velocity $z = x-ct$. Pointwise growth is then relative to the velocity $c$. 
Hence, it is natural to expect for an unstable state that sufficiently large $c$ causes a convective instability, and upon decreasing $c$, a transition to absolute instability can occur at a specific value $c=c_*$. 

Such a $c_*$ is the so-called \emph{linear invasion speed} at which perturbations produce a pointwise impact, and is often the propagation speed of nonlinear fronts which arise from a localized perturbation of an unstable state. 

%travel that result from localised perturbations of a convectively unstable state.  See \S\ref{s:SHE-fronts}.

In order to pursue the ansatz with weighted norms $\|\cdot\|_\eta$, since growth from unstable spectrum is exponential, it is natural to consider exponential weights
\[
\| u \|_{\eta}^2 = \int_\mathbb{R} |e^{-\eta z} u(z)|^2dx,
\]
and the corresponding space $L^2_\eta (\mathbb{R})$ of $u$ with $\|u\|_\eta <\infty$. Notably, the norms for different weights $\eta$ are not equivalent, which is a necessary condition to obtain stability in one space and instability in another. 
In order study a linearisation for perturbations from such a space, we change variables $v(z)=e^{\eta z} u(z)$, which amounts to replacing $\partial_z$ by $\partial_z + \eta$. As a consequence, linear dispersion relations on $L^2(\mathbb{R})$ can be readily transformed to characterise the spectrum on $L^2_\eta(\mathbb{R})$.  

Let us consider the homogeneous steady state $u=0$ and the linearisation in a moving frame $L_c(\mu) := \mu-(1+\partial_z^2)^2 + c\partial_z$ with associated eigenvalue problem  $\lambda u = L_c(\mu) u$ with $\lambda\in \mathbb{C}$. Fourier transforming with Fourier wave number $\ell\in\mathbb{R}$ yields the linear dispersion relation $d_c^*(\lambda,\ell) = \lambda - (\mu - (1-\ell^2)^2 + i c \ell)$, which characterises the spectrum on $L^2(\mathbb{R})$ as $\Sigma_{\mu,c}(0) = \{\lambda\in\mathbb{C} : \exists \ell\in\mathbb{R}, d_c^*(\lambda,\ell) = 0\}$. Notably, stability and instability of the spectrum is independent of $c$; in fact 
$d_c^*(\lambda,\ell) = d_0^*(\lambda - ic\ell,\ell)$. 
For the spectrum on the weighted space $L_\eta^2(\mathbb{R})$, the transformation $\partial_z \to \partial_z + \eta$ amounts to replacing $i\ell$ by $\nu = \eta + i\ell\in\mathbb{C}$. Hence, we obtain the so-called complex (linear) dispersion relation 
\begin{equation}\label{e:cdisp0}
d_c(\lambda,\nu) := \lambda - (\mu - (1 + \nu^2)^2 + c\nu ), \; \lambda,\nu\in \mathbb{C}
\end{equation}
which is the characteristic equation of the linear ODE formed by the eigenvalue problem. 
The spectrum $\Sigma_{\mu,c,\eta}(0)$ for the space $L_\eta^2(\mathbb{R})$ is then 
\[
\Sigma_{\mu,c,\eta}(0) =  \{\lambda\in\mathbb{C} : \exists \ell\in\mathbb{R}, d_c(\lambda,\eta+i\ell) = 0\},
\]
or more explicitly $\lambda = \lambda_{c,\eta}(\ell) := \mu - (1 + (\eta+i \ell)^2)^2 + c(\eta+i\ell )$. The role of $c$ and $\eta$ in the stability properties is rather implicit, but based on the explicit form, much can be understood analytically. While there is no need for numerical continuation to compute $\Sigma_{\mu,c,\eta}(0)$, it can be convenient for the computation of stability boundaries. And for, e.g. reaction diffusion systems, numerical continuation is natural already for the spectrum, where the dispersion relation is the determinant of a matrix, cf.\ \cite{JRSanSch2007}.

Beyond homogenous equilibria, dispersion relations for the spectrum in weighted spaces can be obtained for wavetrains in the same way.  In the Bloch operators \eqref{e:FBop} one again replaces $\partial_\xi$ by $\partial_\xi + \eta$, or alternatively $i\sigma$ by $\nu=\eta + i \sigma$. 
We remark that a wavetrain $u_p$ appears as a temporally periodic orbit for $c\neq 0$ with period $T_c=2\pi/(ck)$ so that its spectral stability properties are characterised by the spectrum of the linearisation of the period map of the abstract solution operator semi-group of \eqref{e:SHeqn}. However, since wavetrains are relative equilibria of the spatio-temporal translation symmetry, spectral stability of the period map is independent of the moving frame. In fact, the spectra in moving frames are related by a shift in the imaginary part of $\lambda$ similar to the case of homogeneous states \cite[Proposition 2.1]{SanSch2000}. Nevertheless, the numerical computation generally requires approximating the time-period map, which is an additional challenge; see for example \cite{dodson2019determining,dodson2024efficient}. 
Concerning spectra of wavetrains in weighted space see also \cite{SanSch2004}.

While spectra in weighted spaces appear to be a natural tool to distinguish absolute and convective instability, on the one hand, it is difficult to identify the appropriate weights and speeds, and the stability boundary of the entire spectrum. On the other hand, these spectra are in general not sufficient since persistent instability need not stem from pointwise growth. See \cite{SanSchabs, dodson2024efficient,trefethen,groot}.  

A robust approach to identify pointwise growth relies on the sign of the real part of certain double roots of the complex dispersion relation \eqref{e:cdisp0}. A general review of this theory and the relation to invasion fronts, along with explicit calculations for the Swift-Hohenberg equation, summarized in \S\ref{s:SHE-fronts}, can be found in \cite{van2003front,holzer2014criteria, goh2023growing}. 

A double root $(\lambda_*,\nu_*)(c)$ of  $d_c$ from \eqref{e:cdisp0} is defined as a solution to
\begin{equation}\label{e:dbleroot}
d_c(\lambda,\nu) = \partial_\nu d_c(\lambda,\nu) = 0.
\end{equation}
Solving $d_c(\lambda,\nu) =0$ for the spatial eigenvalues $\nu$ in terms of $\lambda$ gives four solutions $\nu_j(\lambda)$, $j=1,2,3,4$ that can be chosen to form four continuous curves. At a double root $(\lambda_*,\nu_*)$ we thus have $\nu_j(\lambda_*)=\nu_*$ for (at least) two values $j=j_\pm$. As discussed in the mentioned references, relevant for pointwise growth are ``pinched" double roots $(\lambda_*,\nu_*)$ for which the signs of $\mathrm{Re}(\nu_j(\lambda))$, $j=j_\pm$ are $\pm 1$ as $\mathrm{Re}(\lambda)\to\infty$. It turns out that for \eqref{e:SHeqn} there is a unique complex conjugate pair of double roots $\lambda_*,\bar\lambda_*$ with this property; cf.\  \cite{SRS09}. 

Now the linear spreading speed is determined as 
\begin{equation}\label{e:linspread}
c_* = \sup\{ c\,:\, \mathrm{Re}\, \lambda_*(c)>0\}.
\end{equation}
One can show that $\lambda_*(c_*) = i \omega_*\neq 0$ for some $\omega_*>0$, and $\mathrm{Re} \nu_*(c_*) <0, \quad \mathrm{Im} \nu_*(c_*)\neq0$. Hence, the pointwise growing perturbations oscillate spatio-temporally. As mentioned before, this definition characterises the spreading speed via \emph{marginal stability} of the pinched double root $\lambda_*(c)$. 

Taking an algorithmic and numerical perspective, for a specific value of $\mu$, double roots can be numerically computed algebraically since \eqref{e:dbleroot} is a system of polynomials, but also by a generic Newton method. The dependence of double roots on $\mu$ can be tracked by numerical continuation as discussed before. See Figure~\ref{fig:sh-front} in \S\ref{s:SHE-fronts}. Due to the structure of \eqref{e:SHeqn}, the pinching condition will remain satisfied away from degenerate points of this continuation. In fact, to check the pinching condition it suffices that the other two roots $\nu_j\neq\nu_*$, say $j=1,4$, satisfy $\mathrm{Re}(\nu_1)\leq \mathrm{Re}(\nu_*)\leq \mathrm{Re}(\nu_4)$.

Determining pointwise instability thresholds and the analogue of spreading speeds for wavetrains is more involved. Here it is instructive to briefly discuss the so-called \emph{absolute spectrum} \cite{SanSchabs}, which contains the relevant pinched double roots. 
This set is also interesting as it is the accumulation set of eigenvalues on large bounded domains with separated boundary conditions \cite{SanSchabs}. 

Let us discuss the absolute spectrum for the zero state of \eqref{e:SHeqn}. Its definition is based on $\nu_j(\lambda)$, $j=1,\ldots,4$ as introduced above, ordered by increasing real parts (omitting $\lambda$ for readibility): $\mathrm{Re}(\nu_1)\leq \mathrm{Re}(\nu_2)\leq\mathrm{Re}(\nu_3)\leq\mathrm{Re}(\nu_4)$, while retaining continuity. The absolute spectrum $\Sigma^\mathrm{abs}_{\mu,c}(0)\subset\mathbb{C}$ of the trivial steady state in the moving frame with speed $c$ is then defined as 
\begin{equation}\label{eq:absspec0}
\Sigma^\mathrm{abs}_{\mu,c}(0) := \{\lambda\in\mathbb{C}: \mathrm{Re}(\nu_2)=\mathrm{Re}(\nu_3)\}.
\end{equation}
In order to understand the relation to pinched double roots and the $L^2$-spectrum $\Sigma_\mu(0)$, let $\Omega^*$ be the connected component of $\mathbb{C}\setminus \Sigma_{\mu,c}(0)$ that contains an unbounded part of the positive reals. This set is well-defined since $c$ does not change stability so that $\mathrm{max}\big\{\mathrm{Re}(\Sigma_{\mu,c}(0))\} = \mathrm{max}\big\{\Sigma_{\mu}(0)\} = \mu$; note $\Sigma_{\mu}(0)\subset\mathbb{R}$ is parameterised by $\lambda = \mu - (1-k^2)^2$. On the one hand, in $\Omega^*$ the spatial eigenvalues $\nu_j$ have non-zero real part by definition. On the other hand, scaling $\nu$ as $\mathrm{Re}(\lambda)\to \infty$ readily shows that $\mathrm{Re}(\nu_j)\to -\infty$ for $j=1,2$ and $\mathrm{Re}(\nu_j)\to \infty$ for $j=3,4$. Hence, $\Sigma^\mathrm{abs}_{\mu,c}(0)\cap \Omega*=\emptyset$ and any double root in $\partial\Omega^*$ is a pinched double root. In fact, $\lambda^*, \overline{\lambda^*}$ are the most unstable points (maximal real part) in the absolute spectrum. That the most unstable points are pinched double roots holds more generally, but is not necessary \cite{SRS09, FHSS22}. 
Moreover, the absolute spectrum and the spectrum in weighted spaces, $\Sigma_{\mu,c,\eta}(0)$, are geometrically related\ \cite{JRabs}. In particular,  $\mathrm{max}\big\{\mathrm{Re}(\Sigma_{\mu,c,\eta}(0))\big\}\geq \mathrm{max}\big\{\mathrm{Re}(\Sigma^\mathrm{abs}_{\mu,c}(0))\big\}$, and non-nested self-intersection points of $\Sigma_{\mu,c,\eta}(0)$ lie in $\Sigma^\mathrm{abs}_{\mu,c}(0)$. For some gauge-symmetric systems such as CGL the absolute spectrum actually coincides completely with the weighted spectrum of a specific weight, e.g., \cite{RaSiemer21}.

Up to degenerate points, the absolute spectrum is a union of curves characterised by equal real parts of the spatial eigenvalues with certain index. Hence, locally, the computation can again be done by continuation of solutions to the algebraic set of equations $d_c(\lambda,\nu_j) = 0, j=1,\ldots,4$, $\mathrm{Re}(\nu_2) = \mathrm{Re}(\nu_3)$. Notably, the indices in the ordering by real parts can be determined by computing all four $\nu_j$ and comparing real parts along the continuation. In fact, double roots are natural and convenient initial conditions for a continuation with suitable regularisation, cf.\ \cite{JRSanSch2007}.

This discussion can be translated to the   characterisation of absolute spectrum and pinched double roots for wavetrains. Abstracly, one just needs to replace $\nu_j$ by the Floquet exponents in $\Sigma_{\mathrm{F},0}(\lambda)$ of the period map associated to the spatial ODE for wavetrains, i.e., the Floquet-Bloch spectrum $\Sigma_{\sigma,\tau}(u_p)$ at $\sigma=\tau=0$. Also abstractly, computations of absolute spectrum by numerical continuation can similarly be translated to wavetrains by using the boundary value problem formulation of the Floquet-Bloch spectrum introduced before.
However, it is more challenging to find initial conditions for a continuation; in particular the absolute spectrum for wavetrains is generally not a connected set. This is illustrated for SHE \eqref{e:SHeqn} in Figure~\ref{f:spec}, since absolute and essential spectra coincide for stationary wavertrains. 
Examples where these spectra differ, while retaining the disconnectedness, can be generated by introducing a symmetry breaking term in \eqref{e:SHeqn}, e.g. $\alpha u_{xxx}$, $|\alpha|\ll 1$. The underlying wavetrain is then parameterised by $\alpha$ with $c\neq 0$ for $\alpha
\neq 0$. 

\medskip
Double roots associated to the zero state can be easily computed since $d(\lambda,\nu)$ is a polynomial, but there is no analogue of this for wavetrains, and it is difficult to find a sufficiently good initial guess for a Newton method. It seems that for wavetrains double roots -- and also absolute spectrum -- can only be found from continuation of the weighted spectrum, which itself can be computed from continuation in $\eta$ of the spectrum. As mentioned, good initial conditions for the latter good can be obtained from a discretisation of the eigenvalue problem. See \cite{JRSanSch2007, JRabs}.

\section{Swift-Hohenberg Fronts}\label{s:SHE-fronts}

In this section, we consider pattern-forming invasion front solutions of the Swift-Hohenberg equation. 
It turns out that for the simple supercritical nonlinearity in \eqref{e:SHeqn} the linearized dynamics of the unstable state $u\equiv0$ determines the full nonlinear invasion properties, selecting the invasion speed and as well as the wavenumber of the wavetrain formed in the wake. Such fronts are known as \emph{pulled} fronts. A general review of this theory, along with explicit calculations for the Swift-Hohenberg equation, summarized below, can be found in \cite{van2003front,holzer2014criteria, goh2023growing}, see also \cite{goh2018pattern,KLIKA2024134268,GhaSan2007} for related studies.

The linearization of \eqref{e:SHeqn} in a co-moving frame $z= x - ct$ in $u\equiv 0$ yields the linear operator $L_c(\mu)$ from \S\ref{s:instab} and takes the form 
\begin{equation}\label{e:shl}
v_t = L_c(\mu)v = -(1+\partial_z^2)^2 v + \mu v + cv_z.
\end{equation}
For pulled fronts, the linear predictions from the discussion in \S\ref{s:instab} apply: The pinched double roots $(\lambda_*,\nu_*)(c)$ of the linear dispersion relation $0 = d_c(\nu,\lambda)$ determine the invasion speed $c_*$; the {corresponding} frequency $i\omega_* = \lambda_*(c_*)\neq0$ leads to a temporal oscillation at the leading edge of the front which, through the 1:1 resonance condition $\omega_* = c_*k_*$ (sometimes known as ``node conservation"), determines the wavenumber of the asymptotic pattern as $k_* = \omega_*/c_*$.

We also note that $\nu_*(c_*)$ is an eigenvalue of the spatial dynamical system associated with the linear equation \eqref{e:shl}: 
Scaling time $\taus = \omega t$, 
we consider solutions $v(z,\taus)$ which are $2\pi$-periodic in $\taus$, and with $U = (u,u_z,u+u_{zz},u_z + u_{zzz})^T$ we can write  \eqref{e:shl} formally as the spatial dynamical system 
\begin{equation}\label{e:shl_sd}
U_z = A U, \qquad A = \begin{pmatrix}
0 & 1 & 0 & 0\\
-1 & 0 & 1 & 0\\
0 & 0 & 0 &1\\
\mu - \omega \partial_\taus & c & -1 & 0
\end{pmatrix}.
\end{equation}
On a suitable space of $\taus$-periodic functions, $A$ has compact resolvent, and thus spectrum consisting of only isolated finite-multiplicity point spectrum. 
We remark that the term `spatial dynamics' is also used when (formally) casting the nonlinear \eqref{e:SHeqn} in this form, which is ill-posed as an initial value problem in $z$. For constant functions in $s$ one obtains the nonlinear spatial ODE \eqref{e:SHspatial}, or equivalently \eqref{e:wtBVP}, or \eqref{e:wt-ex}, for the travelling wave profile, that is also often referred to as `spatial dynamics'. 
Equation \eqref{e:shl_sd} turns into the eigenvalue problem of $L_c(\mu)$ when seeking separable solutions $U(z,s) = e^{\lambda s}\tilde U(z)$. Casting \eqref{e:shl_sd} on a space of periodic functions only purely imaginary such eigenvalues appear. Since we seek invasion fronts as periodic functions in $s$, this formulation is suitable in the present context. 
One can study the spectrum of $A$ by decomposing in Fourier series, $U = \sum_j e^{i j \taus} \hat U_j$, finding that $\nu_*, \bar\nu_*$ are algebraically-double and geometrically simple eigenvalues of $A$, in the $j = \pm 1$ Fourier subspaces respectively (each of which are invariant under the flow of \eqref{e:shl_sd}),  with eigenfunction $U_j = (1,\nu_j,(1+\nu_j^2),\nu_j(1+\nu_j^2) ).$

After scaling the co-moving frame variable $z$ by the asymptotic wavenumber parameter $k$, we seek modulated front solutions of the form
$$
u(x,t) = u_\mathrm{f}(\zeta,\taus), \qquad \zeta =  kz =  k(x - ct),\,\, \taus = \omega t, 
$$
which connect the trivial state to the periodic state $u_p(k x;k)$ as $\zeta\to\pm\infty$. Recall $u_p$ satisfies \eqref{e:wtBVP}-\eqref{e:wt_phase}, introduced in \S\ref{s:wtonset}.

Wavetrains $u_p$ in the Swift-Hohenberg equation are stationary in a stationary frame and thus are periodic in a co-moving frame, satisfying $u_p(k x;k) = u_p(\zeta + k c t;k)$. We look for fronts which are 1:1 resonant with the asymptotic wavetrain with wavenumber $k$, namely  $\omega = k c$ so that $u_p(k x;k) = u_p(\zeta + \taus;k)$d. In sum we obtain the following traveling wave equation,
\begin{equation}\label{e:shmtw}
0 = -(1+k^2\partial_\zeta^2)^2 u_\mathrm{f} + \mu u_\mathrm{f} - u_\mathrm{f}^3 + c k(u_{\mathrm{f},\zeta} - u_{\mathrm{f},\taus}), \quad\zeta\in \mathbb{R}, \,\, \taus\in (0,2\pi]. 
\end{equation}
We impose the following asymptotic boundary conditions to study pattern forming front solutions
\begin{equation}\label{e:shbc}
\lim_{\zeta\rightarrow-\infty}|u_\mathrm{f}(\zeta,\taus) - u_p(\zeta + \taus;k)| = 0, \qquad 
\lim_{\zeta\rightarrow\infty} u_\mathrm{f}(\zeta,\taus) = 0.
\end{equation}

Due to the moving frame and rescaling \eqref{e:shl} can be written as  
\[
\mathbb{L} v=0, \quad 
\mathbb{L}:=  (1+k^2\partial_\zeta^2)^2 + \mu + c k(\partial_\zeta - \partial \taus).
\]
In scaled coordinates, we then have
\begin{equation}\label{e:Lnu}
\mathbb{L} 
\left( (\alpha \zeta + \beta)e^{\tilde \nu_* \zeta + \taus}  \right) = 0,
\end{equation}
for arbitrary $\alpha,\beta\in \mathbb{C}$, and $\tilde\nu_* = \nu_*/k$ to account for the scaling from $z$ to $\zeta$. 
%For ease of notation we also let $N(u) = -u^3$. 

To localize the front solution in space, we use a farfield-core decomposition. This technique more generally separates the far-field behavior -- which often is known explicitly or satisfies a less-complicated equation -- from the interfacial behavior connecting the two asymptotic states.  It has been successfully used in both theoretical and numerical studies of general spatio-temporally heteroclinic phenmomena \cite{lloyd2019invasion, lloyd2017continuation, dodson2019determining,morrissey2015characterizing}. Spatial localization of the nonlinear problem helps regain Fredholm properties of the associated linearization, mitigate neutral continuous spectrum, and even singular perturbations. 

\begin{figure}[ht]
    \centering
    \hspace{-0.5in}\includegraphics[width=0.48\linewidth]{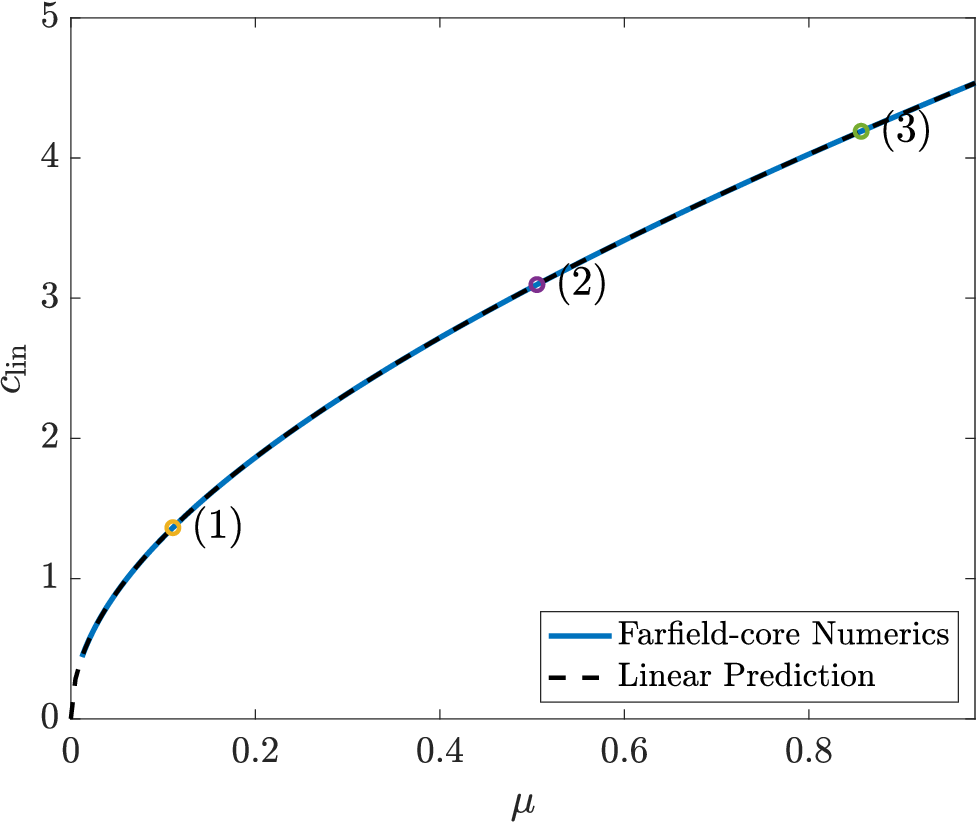}\hspace{-0.05in}
    \includegraphics[width=0.5\linewidth]{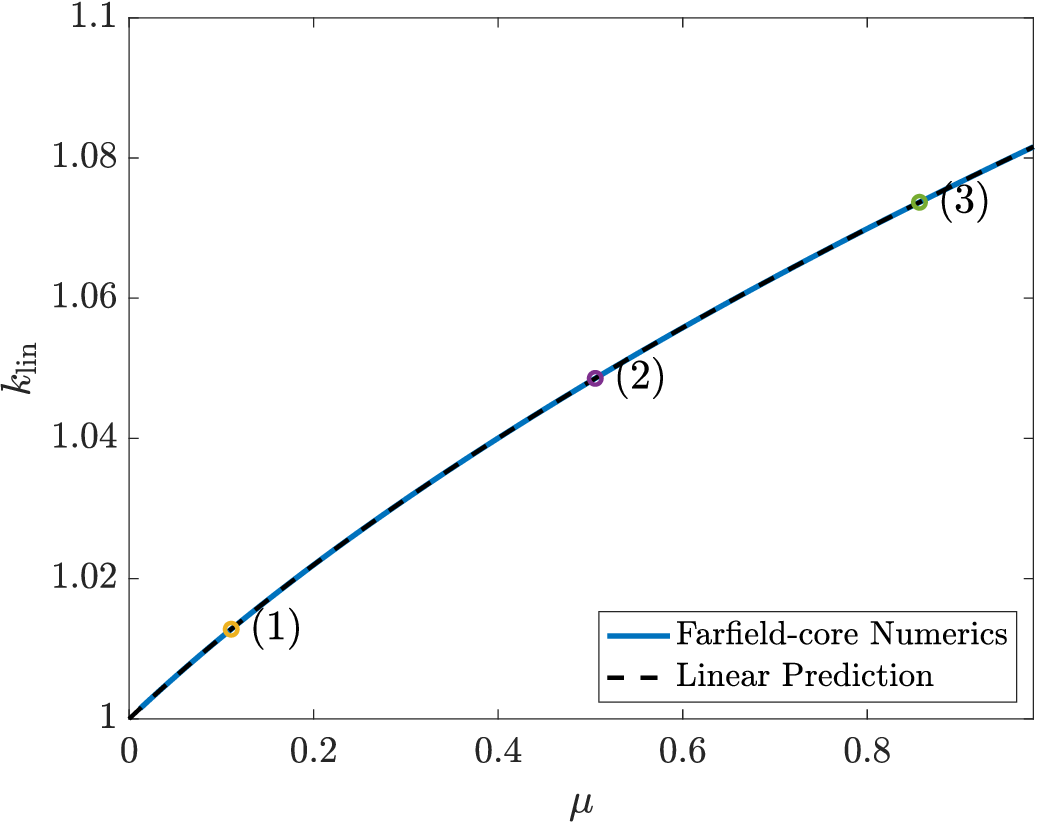}\hspace{-0.05in}\\
      \hspace{-0.5in}\includegraphics[width=0.5\linewidth]{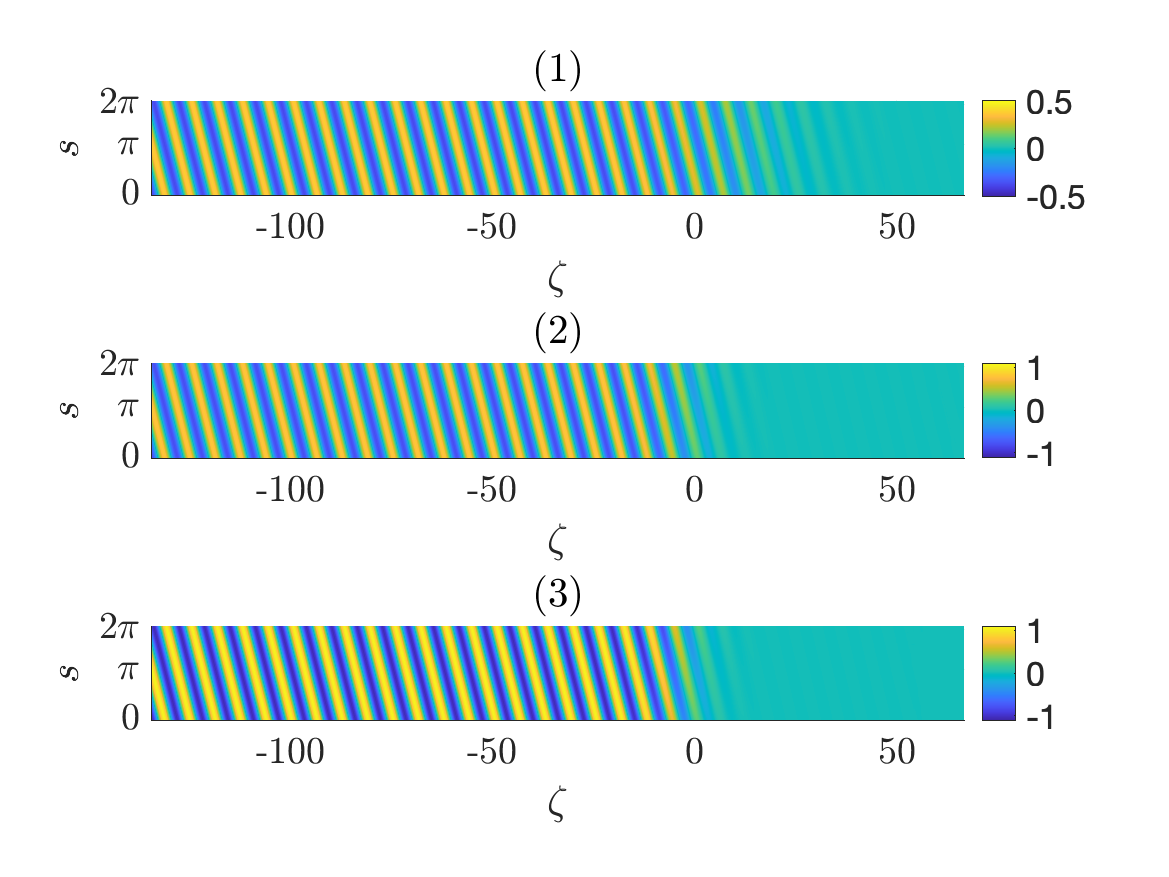}\hspace{-0.15in}\hspace{-0.05in}
        \includegraphics[width=0.5\linewidth]{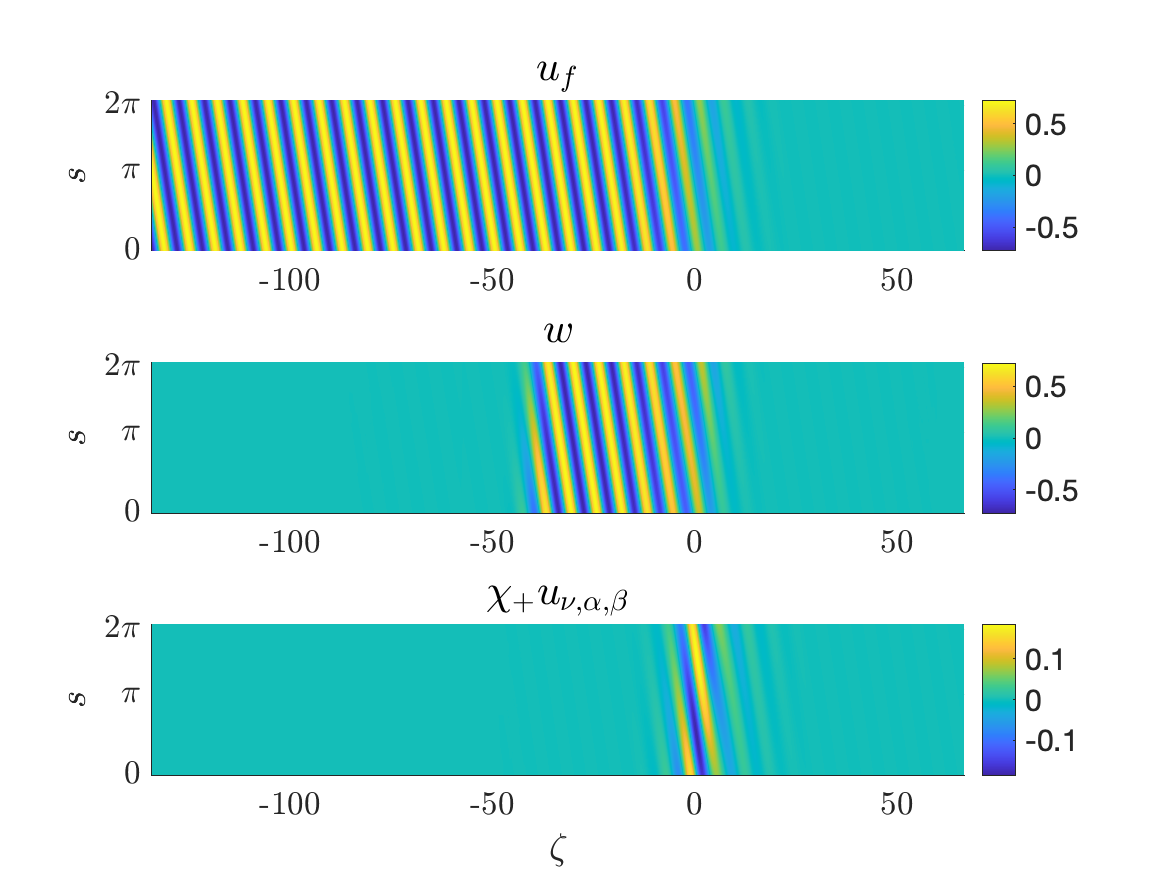}\hspace{-0.3in}
    \caption{ Upper Left: comparison of numerically continued invasion speed (solid blue) compared against linear prediction (dashed black) for $c_*$; Upper right: comparison of numerical continuation (solid blue) and linear prediction (dashed black) for $k_*$; Lower left: Front solutions $u_\mathrm{f}$  of \eqref{e:shcont1} - \eqref{e:phcon} for three values of $\mu$, labeled in left and center figures; Lower right: Depiction of full front solution (top) along with core solution $w$ (middle) and leading edge far-field term $\chi_+(\zeta)(\alpha \zeta + \beta)e^{\nu \zeta/k + i \taus}$ for $\mu = 0.28132.$}
    \label{fig:sh-front}
\end{figure}

We follow the approach for pulled invasion fronts described by \cite{avery2023pushed} and define 
$$
u_{\tilde\nu,\alpha,\beta}(\zeta,\taus):= (\alpha \zeta + \beta) e^{\tilde\nu\zeta + i\taus}+\mathrm{c.c.},
$$ 
as well as the far field-core ansatz 
\begin{equation}
\label{e:cffa}
u_\mathrm{f}(\zeta,\taus) = \chi_-(\zeta) u_p(\zeta +  \taus;k)+
w(\zeta,\taus) 
+\chi_+(\zeta) u_{\tilde\nu,\alpha,\beta}(\zeta,\taus). %(\alpha \zeta + \beta) e^{\tilde\nu\zeta + i\taus}.
\end{equation}
Here $\chi_+(\zeta) = (1+e^{(\zeta-1)/m})^{-1}$ and $\chi_-(\zeta) = (1+e^{-(\zeta - \zeta_0)/m})^{-1} $ are cutoff functions, nearly 1 on the right and left hand side of the spatial domain respectively, where $m>0$ and $\zeta_0<0$ are fixed constants; see Figure \ref{fig:sh-front} (lower right) for a depiction of this decomposition. We call the $2\pi$-periodic function  $w(\zeta,\taus)$  the core part of the solution. It resolves the interface between the two far-field states which are inserted explicitly into the decomposition using the cutoff functions. 
On the left of the spatial domain, we insert the wavetrain solution $u_p$ and control it via the parameter $k$. On the right, to account for the partially algebraic convergence and to further localize $w$, we insert the expected asymptotic tail of the front determined by the linearized dynamics. We control this solution through the Jordan block parameters $\alpha,\beta$ and the scaled spatial eigenvalue parameter $\tilde \nu = \nu/k$ which solves the (scaled) dispersion relation \eqref{e:dbleroot}. We then expect the core function $w$ to be more strongly localized in $\zeta$,  with decay rate stronger than $\mathrm{Re} \tilde \nu$ at $\zeta\gg1.$ This localization is generally caused by the exponential convergence in $\zeta$ of the front to its asymptotic states which itself is due to the hyperbolicity of the two far-field states; see \cite{avery2023pushed,goh2018pattern,goh2023growing} for similar arguments in a slightly different context.

Note here we have made $\tilde \nu$, $\alpha$,$\beta$, $k$, and $c$ free variables to be solved for by the continuation algorithm at each step. 
Inserting the ansatz \eqref{e:cffa} into \eqref{e:shmtw} and subtracting off and using the following identities
\begin{align}
0 &= \chi_+(\zeta) \mathbb{L}[u_{\tilde\nu,\alpha,\beta}(\zeta,\taus) ] \notag\\
0 &=  \chi_-(\zeta) [\mathbb{L} u_p(\zeta+\taus;k) -u_p(\zeta+\taus;k)^3] %+ N(u_p(\zeta+\taus;k)],%[ -(1+k^2\partial_\zeta^2)^2u_p +ck\partial_\taus u+ \mu u_p +N(u_p)].
\end{align}
we obtain the equation
\begin{align}\label{e:shcont1}
    0 &= \mathbb{L} w + c k(w_\zeta -w_\taus)  
%+ N(w+\chi_+u_{\tilde\nu,\alpha,\beta} +\chi_- u_p(\cdot;k)) 
-(w+\chi_+u_{\tilde\nu,\alpha,\beta} +\chi_- u_p(\cdot;k))^3
+  g,%(\zeta,c,k,\tilde \nu,\alpha,\beta),\quad 
\end{align}
where 
$$
g=g(\zeta,c,k,\tilde \nu,\alpha,\beta) = [\mathbb{L},\chi_+]u_{\tilde \nu,\alpha,\beta} + [\mathbb{L},\chi_-]u_p +\chi_- u_p^3%N(u_p)
$$
and $[\mathbb{L},\chi_-]v = \mathbb{L}(\chi_- v) - \chi_-\mathbb{L}v$ denotes the commutator between $\mathbb{L}$ and $\chi_-$.  
We remark that the above subtraction of the far-field terms takes advantage of the fact that the asymptotic states solve a known equation and reduces round-off errors in the farfield in numerical computations. 
Due to the spatial and temporal translational symmetries, phase conditions are required to obtain local uniqueness for solutions of \eqref{e:shcont1}. 
In sum, modulo such translations, front solutions $u_\mathrm{f}$ can be characterized by the core-variable $w$, the far-field parameters $\tilde\nu,\alpha,\beta,$ the front speed $c$, and the system parameter $\mu$. 

To numerically approximate such front solutions, we truncate this asymptotic boundary value problem for $w$ onto a finite domain in $\zeta$, %and of course 
retaining the periodicity of the domain in $\taus$. Since we expect that $w$ will be strongly exponentially localized in $\zeta$ (i.e. exponentially close to zero near the boundary) %one can 
we could impose both Dirichlet or periodic boundary conditions in $\zeta$.  Due to the asymptotic hyperbolicity of the asymptotic states of the front, we expect the truncation error to be exponentially small in $L_\zeta\gg1$. We note that there are no 
rigorous convergence results  
%results rigorously convergence resultsd 
connecting solutions of the truncated boundary value problem with the unbounded domain front solution; see \cite{jankovic} for some preliminary results in this direction.  As we wish to take advantage of the computational efficiency and accuracy of spectral methods and the Fast Fourier Transform, we impose periodic boundary conditions in $\zeta$.   

Our computational domain will be $(\zeta,\taus)\in I_\zeta\times (0, 2\pi]$ for $I_\zeta:=(-5L_\zeta/3,L_\zeta/3]$. {Here $5/3$ is a somewhat arbitrary choice, giving a} computational domain shifted in $\zeta$ since the subtraction of the leading-edge of the front and strongly localized $w$, allows a smaller domain on the right hand side. 

In order to compute and continue the pulled fronts, we follow the approach suggested in the recent work \cite{avery2023pushed}.
We discretize the equation pseudo-spectrally, taking advantage of the parallelizability of the Fast Fourier Transform in MATLAB, with a total of $N = N_\zeta N_\taus$ Fourier modes. We let $W = \{W_{i,j}\}_{i = 1...N_\zeta,j = 1...N_\taus}$ denote the discretized core function on the corresponding spatial mesh. We append the following equations to \eqref{e:shcont1} in order to solve for all of the free variables,
\begin{align}
0 &= d(\nu,i c k, c),\qquad 0 = \partial_\nu d(\nu,i c k, c),\label{e:shdd}\\
%0&= \sum_{i = 1}^{N_\taus} W_{N_\zeta,i} + W_{N_\zeta-1,i},\\
0&=  W_{N_\zeta,N_\taus/2} + W_{N_\zeta-1,N_\taus/2},\label{e:tail1}\\
0&=  \int_{0}^{2\pi}\int_{d-2\pi}^{d} w(\zeta,\taus) \cos(\taus) d\zeta d\taus \label{e:tail2}\\
0&= \int_{0}^{2\pi}\int_{I_\zeta} e^{-\zeta^2} u_\mathrm{f}(\zeta,\taus) d\zeta d\taus.\label{e:ph-uf} \\
0&= \int_{0}^{2\pi}\int_{-d-2\pi}^{-d} u_p'(\zeta + \taus;k) w(\zeta,\taus) d\zeta d\taus, \label{e:phcon}
    \end{align}
for some $d\in (0,L_\zeta/4)^T$ ($d = 5\pi/2$ in the present computations
%\JRx{Why this choice?}\RGx{To be honest, it was what worked after a large amount of experiementation}
). 
The first two conditions in \eqref{e:shdd} solve for the double root $(\nu,\lambda)$ in the unscaled $\xi$ coordinates, the third and 4th equations \eqref{e:tail1}, \eqref{e:tail2} are transversality conditions requiring the tail of $w$ at $\xi\gg1$ to decay sufficiently fast, i.e. lie outside the eigenspaces $\mathrm{span}\, U_j,\,\, j = 0,\pm 1$ described above, while the last two equations \eqref{e:ph-uf}, \eqref{e:phcon} are phase conditions which fix translations of the front in $\zeta$ and $\taus$ respectively. 

Hence we have $N+8$ equations (counting each complex equation as two real ones written in terms of the real and imaginary parts of each variable), and $N+8$ real variables $\tilde W = (W,\nu = \nu_r+i\nu_i,c,k,\alpha = \alpha_r + i \alpha_i,\beta = \beta_r + i\beta_i)$. We append $\mu$ as the main the continuation parameter and use a secant continuation approach, 
in particular using a standard pseudo-arclength continuation condition $0 = \int_{I_\zeta\times[0,2\pi)} (w-w_{old})^T w_\mathrm{sec}\, d\zeta d\taus$, where $w_\mathrm{sec}$ is the normalized secant vector given by the previous two continuation steps.

We use a Newton-GMRES approach to solve each step, with the Jacobian of the above discretized set of equations given to MATLAB's GMRES algorithm as a function at each Newton step. Each linear solve is preconditioned by the (spectrally formed) Fourier multiplier operator $P = (-1+\mathbb{L})^{-1}$. Derivatives in the parameter variables $(\alpha_r,\alpha_i,\beta_r,\beta_i,\nu_r,\nu_i, k,c,\mu)$ are calculated using simple finite difference evaluations of the nonlinear system. The auxiliary small domain problem for $u_p$, called in every evaluation of the full nonlinear function, is discretized on $\theta\in [0,2\pi)$ with 16 modes and solved in a similar manner to the larger problem.  This small-domain periodic solution is then periodically extended (in spectral space) onto the large domain core problem. See \cite{goh2023growing} for more details of this type of approach on a different, but related problem. 

%Unsuprisingly, as the algebraic equations determining the linear spreading speed/wavenumber are explicitly included in the computation, the results of our numerical continuation of the discretized nonlinear system agree quite well with the linear predictions given above.  
As expected from the theoretical predictions, the results of our numerical continuation of the discretized nonlinear system agree quite well with the linear predictions given above.

Figure \ref{fig:sh-front} (top row) gives a comparison of the numerically continued front speed and wavenumber with the linear predictions. Figure \ref{fig:sh-front} (lower left) depicts the front solution for a few select $\mu$ values (labeled on the curves in the top row). Note that the asymptotic amplitude of the periodic pattern $u_p$ varies in $\mu$, with scaling $\sim \mu^{1/2}$ for $\mu\sim 0$; see \eqref{e:wavetrain} above.   Figure \ref{fig:sh-front} (lower right) also depicts the different compuational components, including the full front $u_\mathrm{f}$, the core function $w$ and the far-field term for the right leading edge of the front,  $\chi_+u_{\tilde\nu,\alpha,\beta}$. We find, as expected, the core function $w$ is localized in between the supports of the cut-off functions $\chi_\pm$.

 We remark that such a computational approach is generalizeable to patterned invasion in higher-order scalar and multi-component PDE models, including those where the dispersion relation is too complicated and does not yield explicit formulas for the spreading speed and selected wavenumber. We also remark that this approach lends itself naturally to massive parallelization using multi-core CPUs as well as GPUs. Indeed, the code used to obtain the above results can be readily altered to run on a GPU using the NVIDIA CUDA wrappers built into MATLAB.

%%%%%%%%%%%%%%%%%%%%%%%%%%%%%
\section{Discussion}\label{s:discus}
In writing this chapter, we envisioned a numerical continuation tutorial for the very classical question: How fast does a wavetrain invade an unstable equilibrium in the Swift-Hohenberg prototypical pattern forming system? However, it is surprisingly difficult to locate in the literature numerical investigations {of even the more basic}  existence and stability of wavetrains for the SHE \eqref{e:SHeqn}  away from small $|\mu|$, similar to Figure~\ref{f:SHE-Busse}. 
The intricate existence boundary of the Turing instability, shown in Figure~\ref{f:wt-existence} for $\mu=0.9$ (i.e. before the bifurcation of the additional equilibria), has yet to be fully explored as far as we are aware though there have been some investigations in this direction \cite{Peletier2004,Peletier2007}. We again refer to \cite{JB2008} for insights into the somewhat abstract complexity of steady states. 
The relevance of our central question has recently been looked at via marginal stability and Ginzburg-Landau equation analysis for reaction-diffusion equations~\cite{Klika82024} with an application to biological systems and model selection, demonstrating that this is a vibrant research area. Indeed, the numerical results we  present go beyond a review and we encourage readers to explore this further. For instance, to our knowledge the stability of the coherent invasion front has, in general, not been corroborated beyond modulational analysis and direct numerical simulations. {In the Fitzhugh-Nagumo system, a stability study has been conducted recently for the specific case of a rigidly propagating front, where the phase velocity of stripes is locked with propagation speed  \cite{avery2023stabilitycoherentpatternformation}.}

All the methods detailed here can be readily extended to more general systems, such as the reaction-diffusion set up; as mentioned our use of the plain cubic SHE \eqref{e:SHeqn} is for illustrative purposes. For the 1D stability analysis of Turing patterns and numerical continuation for general reaction-diffusion systems, an implementation in AUTO~\cite{auto} has been outlined in \cite{JRSanSch2007}. It has been framed into the package WAVETRAIN~\cite{Sherratt2012}, which relies on AUTO07p to carry out the computation and numerical continuation of the spatially periodic orbits and Eckhaus instability boundaries. This can readily be extended to 2D stability, in particular zizgag boundaries as described in \S\ref{s:zigzagnum}. The software PDE2PATH~\cite{p2pbook} is able to compute and continue wavetrains in more general PDE systems and 2D (even 3D), and to calculate co-periodic stability. There appears to be a gap in the  openly available software for the zigzag stability analysis for general systems. 

The use of exponential weights to compute the spectra has been outlined and illustrated in \cite{JRSanSch2007,JRabs}; %but largely only been done on an ad-hoc basis, and 
setting this up in a general software package would be highly useful for the community, and is planned for a future release of PDE2PATH. However, we hope the interested reader can use 
%the methods described in this chapter can be readily 
this chapter for an implementation in their favorite continuation software such as COCO (see previous chapter in this book), \texttt{BifurcationKit}~\cite{bifkit},  MATCONT \cite{matcont}, or based on our implementation provided under the github link in \S\ref{s:intro}.  
We note that the continuation approach can also be applied to other destablisation mechanisms, in particular transverse instabilities for striped patterns in 2D, and to other types of patterns such as hexagons (see for instance~\cite{Lloyd2021} in the nonlinear selecting hexagon invasion front setting) or travelling defects~\cite{SanSch2004}. 

The far-field core decomposition numerics carried out in \S\ref{s:SHE-fronts}, in principle, can be extended to general reaction-diffusion systems. However, as detailed in \S\ref{s:SHE-fronts}, the numerics are very sensitive to the choice of computational parameters and requires spatial dynamical systems knowledge to implement, hence developing robust numerical routines remains an area for further research. 
The power of the far-field core decomposition approach has been demonstrated in a variety of different contexts where one is interested in computing solutions which connect to non-trivial asymptotic behaviour with unknown parameters which have to be solved for as part of the connection problem. 
For instance, the far-field core decomposition has been used to compute 2D grain boundaries~\cite{lloyd2017continuation}, stripes on the half-line with non-trivial boundary conditions at the origin~\cite{Morrissey2015}, spectra of spirals~\cite{dodson2024efficient,dodson2019determining}. See \S\ref{s:SHE-fronts} for a few further references. 
We also note the recent data-driven numerical continuation approach~\cite{Zhao2024} that could be a potentially interesting avenue to explore and extend to the context considered in this chapter.

\bibliographystyle{plainurl}
\bibliography{ch_GLR_arxiv}
\end{document}